\documentclass[prd, 
 reprint,
 superscriptaddress,
preprintnumbers,
nofootinbib,
 amsmath,
 amssymb,
 aps,
]{revtex4-2}
\usepackage{gensymb}
\usepackage{graphicx}
\usepackage{dcolumn}
\usepackage{bm}
\usepackage{hyperref}
\usepackage[export]{adjustbox}
\usepackage{subfiles}
\usepackage{booktabs}

\usepackage{siunitx}
\DeclareSIUnit[]\Msolar
{\text{\ensuremath{M_{\odot}}}}
\DeclareSIUnit \parsec {pc}

\usepackage[acronym]{glossaries}

\newacronym{gw}{GW}{gravitational wave}
\newacronym{lvk}{LVK}{LIGO-Virgo-KAGRA}
\newacronym{snr}{SNR}{signal to noise ratio}
\newacronym{eob}{EOB}{effective one body}
\newacronym{imr}{IMR}{Inspiral--Merger--Ringdown}
\newacronym{svd}{SVD}{singular value decomposition}
\newacronym{psd}{PSD}{power spectral density}
\newacronym{bh}{BH}{black hole}
\newacronym{ns}{NS}{neutron star}
\newacronym{bbh}{BBH}{binary black hole}
\newacronym{bns}{BNS}{binary neutron star}
\newacronym{isco}{ISCO}{innermost stable circular orbit}
\newacronym{nsbh}{NSBH}{neutron star -- black hole}
\newacronym{bhns}{BHNS}{Black Hole -- Neutron Star}
\newacronym{pdf}{PDF}{probability density function}
\glsdisablehyper

\begin{document}

\title{Identifying Eccentricity in Binary Black Hole mergers using a Harmonic Decomposition of the Gravitational Waveform}

\author{Ben G. Patterson}
\email{pattersonb1@cardiff.ac.uk}
\affiliation{%
Gravity Exploration Institute, School of Physics and Astronomy, Cardiff University, Cardiff, CF24 3AA, United Kingdom
}

\author{Sharon Mary Tomson}
\affiliation{%
Gravity Exploration Institute, School of Physics and Astronomy, Cardiff University, Cardiff, CF24 3AA, United Kingdom
}
\affiliation{%
Max-Planck-Institut für Gravitationsphysik (Albert-Einstein-Institut), Callinstraße 38, D-30167, Hannover, Germany
}
\affiliation{%
Leibniz Universität Hannover, D-30167, Hannover, Germany
}

\author{Stephen Fairhurst}
\affiliation{%
Gravity Exploration Institute, School of Physics and Astronomy, Cardiff University, Cardiff, CF24 3AA, United Kingdom
}

\begin{abstract}
We show that the gravitational waveform emitted by a binary on an eccentric orbit can be naturally decomposed into a series of harmonics.  The frequencies of these harmonics depend upon the radial frequency, $f_{\mathrm{r}}$, determined by the time to return to apoapsis, and the azimuthal frequency, $f_{\phi}$, determined by the time to complete one orbit relative to a fixed axis.  These frequencies differ due to periapsis advance. Restricting to the (2, 2) multipole, we find that the frequencies can be expressed as $f = 2 f_{\phi} + k f_{\mathrm{r}}$. 
We introduce a straightforward method of generating these harmonics and show that the majority of the signal power is contained in the $k= -1, 0, 1$ harmonics for moderate eccentricities.  We demonstrate that by filtering these three leading harmonics, we are able to obtain a good estimate of the orbital eccentricity from their relative amplitudes.
\end{abstract}

\maketitle

\section{Introduction}
\label{Sec:Intro}

\Gls{gw} data analyses commonly assume observed GW signals to originate in quasi-circular binary systems with a negligible amount of eccentricity \cite{KAGRA:2021vkt}. This is motivated by the general relativity prediction that compact binaries rapidly circularise during inspiral \cite{Peters:1963ux, Peters:1964zz} with the value of eccentricity approximately halving when the orbital frequency is doubled in the low eccentricity limit \cite{Favata:2021vhw, Fumagalli:2024gko}. Studies have shown that matched filtering eccentric signals with quasi-circular templates can cause significant power loss \cite{Martel:1999tm, Brown:2009ng, Barack:2003fp, Gadre:2024ndy}, potentially pushing observable \gls{gw} events under the detection threshold. For eccentric signals that are detected, it can be shown that performing parameter estimation with quasi-circular waveforms introduces systematic biases in the recovered posteriors \cite{Barack:2003fp, Favata:2021vhw, Gadre:2024ndy, Gupte:2024jfe}. The inclusion of eccentricity is therefore crucial to the detection and accurate analysis of eccentric \gls{gw} events. 

For a field binary of massive stars that evolve into \glspl{bh} at the end of their lives, we expect the \gls{bbh} to radiate away any eccentricity before entering the detectable band of current \gls{gw} detectors \cite{Moore:2016qxz, Fumagalli:2024gko}. While eccentricity can be introduced to systems in this isolated formation channel, for example through stellar kicks from the second supernova \cite{Saini:2023wdk}, it is expected that the majority of observed \gls{gw} signals with non-negligible eccentricity will be from \glspl{bbh} that have formed dynamically \cite{Favata:2021vhw, Fumagalli:2024gko, Zevin:2021rtf, Vijaykumar:2024piy}. \Glspl{bh} in dynamical environments such as globular clusters or galactic nuclei can form gravitationally bound eccentric binary systems with relatively low orbital separations, causing the \glspl{bh} to merge before all of the eccentricity can be radiated away. In these cases the \gls{bbh} may still have a significant value of eccentricity at detectable frequencies \cite{OLeary:2005vqo, Favata:2021vhw, Fumagalli:2024gko, Zevin:2021rtf}. The detection (or non-detection) of eccentricity in \gls{gw} signals can thus place constraints on origins of observed \gls{bbh} systems \cite{Zevin:2021rtf}.

Traditional Bayesian parameter estimation techniques used for quasi-circular signals are extremely computationally expensive when applied to eccentric waveforms for two main reasons. Firstly, the detailed effects introduced by eccentricity cause waveforms with non-zero eccentricity to be much slower to generate \cite{Ramos-Buades:2023yhy}. Secondly, two additional parameters must be added to the analysis: the eccentricity and the mean anomaly, which determines the position along the eccentric orbit \cite{Ramos-Buades:2023yhy, Clarke:2022fma}. Several studies have analysed a small selection of events using this approach \cite{Ramos-Buades:2023yhy, Bonino:2022hkj, Gamba:2021gap}, however performing full Bayesian analyses including eccentricity on a catalogue of \gls{gw} events remains computationally challenging. With the higher event rates and detector sensitivity of O4 and O5 \cite{KAGRA:2013rdx}, there is a growing need for faster parameter estimation techniques incorporating eccentricity to keep up the rate of observations. 

Several studies have developed and applied such techniques to search for evidence of eccentricity in the first three \gls{lvk} \cite{LIGOScientific:2014pky, VIRGO:2014yos, KAGRA:2020tym} observing runs \cite{KAGRA:2021vkt, LIGOScientific:2019dag, LIGOScientific:2023lpe}. Romero-Shaw et al.~\cite{Romero-Shaw:2019itr, Romero-Shaw:2022xko} reweight samples from quasi-circular analyses and find evidence for eccentricity in GW190521, GW190620, GW191109, and GW200208\_22. Gupte et al.~\cite{Gupte:2024jfe} have more recently used the machine-learning code \texttt{DINGO} to accelerate parameter estimation and find support for eccentricity in GW190701, GW200129, and GW200208\_22. Nevertheless, there is no universally agreed GW candidate with a confirmed detection of non-zero eccentricity. A major challenge both of these studies have faced is distinguishing between the effects of eccentricity and precession. It is widely believed that an observational degeneracy exists between the two~\cite{Romero-Shaw:2022fbf, Xu:2022zza}, and work is underway to produce a waveform approximant with functionality to simultaneously model both \cite{Liu:2023ldr}.

In this paper we propose a method based on decomposing the eccentric waveform into harmonics. Such a decomposition has been described in many previous studies, both in terms of half-integer multiples of the standard circular \gls{gw} frequency\cite{Peters:1964zz, Moore:2018kvz}, and by including a new frequency induced by apsidal advance \cite{Moreno:1995, Seto:2001pg, Willems:2007nq, Valsecchi_2012, Moore:2016qxz}. \Gls{svd} has been used in the context of \glspl{gw} before improve the efficiency of searches \cite{Cannon:2010qh}. Here we apply \gls{svd} in order to identify the most efficient harmonic decomposition of eccentric waveforms and analyse its structure.

The amplitude of eccentric harmonics
depends on the eccentricity of the system \cite{Peters:1963ux, Moreno:1995, Seto:2001pg}. By calculating the power in two or more harmonics in a real \gls{gw} signal, we can then compute a quick estimate of the binary's eccentricity. A similar approach has been applied in the past for higher order multipoles and precession harmonics~\cite{Fairhurst:2023idl}. Developing a physical understanding of eccentric harmonics may also allow us to compare their form to precession harmonics and shed light on the cause and extent of any degeneracy that exists between the two.

In Sec.~\ref{Sec:EccWaveform} we describe the main features of eccentric waveforms and identify the frequencies of their harmonic structure. In Sec.~\ref{Sec:Decomposition} we apply \gls{svd} to a set of eccentric waveforms to identify a fast and robust way to generate eccentric harmonic waveforms from existing waveform models. We examine the degeneracy between eccentricity and chirp mass in Sec.~\ref{Sec:Degeneracy} and show how we can use quasi-circular parameter estimation to inform the starting point of our method. In Sec.~\ref{Sec:Ecc_dist} we describe how to map from calculated harmonic \glspl{snr} to eccentricity, and apply this to a simulated example.  Finally, in Sec.~\ref{Sec:Discussion}, we provide a summary and discussion of future applications of our approach.

\section{Gravitational Waveforms from Eccentric Binaries}
\label{Sec:EccWaveform}

The calculation of full binary merger waveforms for eccentric signals is challenging.  The first extension to the leading order calculation \cite{Peters:1963ux, Peters:1964zz} was to include post-Newtonian effects \cite{Junker:1992kle, Blanchet:1989cri, Gopakumar:1997bs, Memmesheimer:2004cv, Arun:2007rg, Arun:2007sg} which provide a more accurate description of the waveform during the inspiral phase. Full numerical simulations of  binaries on eccentric orbits have been performed and compared to post-Newtonian waveforms \cite{Hinder:2008kv}.  More recently, complete waveform models, combining information from post-Newtonian theory and numerical simulations have been developed \cite{Hinder:2017sxy, Islam:2021mha, Nagar:2021gss, Ramos-Buades:2021adz}. There exist several eccentric waveform models capable of use in searches for and parameter estimation of non-circular \gls{bbh} systems \cite{Chiaramello:2020ehz, Nagar:2021gss, Nagar:2024dzj, Ramos-Buades:2021adz, Liu:2023dgl, Paul:2024ujx, Tanay:2016zog}. In this work we choose to use the \texttt{TEOBResumS-Dali} waveform model in our analysis, however the method and techniques described can be easily expanded to other eccentric waveforms, taking care to consider differing definitions of eccentricity used by each waveform model~\cite{Knee:2022hth, Vijaykumar:2024piy}. A definition of eccentricity, $e_\mathrm{gw}$, has been proposed \cite{Shaikh:2023ypz} which measures the eccentricity directly from the waveform after it has been generated.  However we leave the incorporation of this to future work, and simply use the value of eccentricity as reported by the \texttt{TEOBResumS-Dali} waveform model. \texttt{TEOBResumS-Dali} uses the \gls{eob} \cite{Buonanno:1998gg} formalism to construct full models of the \gls{imr} waveform. This model incorporates the effect of aligned spins on the emitted waveform but does not include in-plane spins which lead to orbital precession \cite{Apostolatos:1994mx}.

\subsection{Waveforms from Eccentric Binaries}
\label{SubSec:EccWFs}

The leading order gravitational wave emission from an eccentricity binary system was first calculated in \cite{Peters:1964zz, Peters:1963ux}.  As is well known, the orbital eccentricity leads to an increase in gravitational wave emission, relative to a circular binary, and a decay of eccentricity as the orbit shrinks.  
In addition, relativistic effects lead to advance of the periapsis \cite{Einstein:1915bz}.  Both of these effects impact the nature of the gravitational wave signal emitted by a binary on an eccentric orbit.  Here, we briefly recap the key features of the waveform.  We refer readers to other papers, e.g.~\cite{Peters:1963ux, Moreno:1995, Arun:2007sg, Moore:2016qxz}, for more complete descriptions of the evolution of eccentric binaries.

Consider a binary with masses $m_{1}$ and $m_{2}$ on an elliptical orbit with a semi-major axis $a$ and eccentricity $e$.  As with a circular binary, the evolution is determined, at leading order by the chirp mass, $\mathcal{M}$, defined by
\begin{equation}\label{Eq:chirp_mass}
    \mathcal{M} = \frac{(m_{1} m_{2})^{3/5}}{(m_{1} + m_{2})^{1/5}} \, ,
\end{equation} 
and consequently we choose to parametrize the system by the chirp mass and mass ratio, defined as
\begin{equation}\label{Eq:mass_ratio}
    q = \frac{m_2}{m_1} 
    \quad \mbox{where} \quad 
    m_{2} \le m_{1} \, .
\end{equation}
The orientation of the binary at a given time $t$ is given by the orbital phase $\phi(t)$ relative to a fixed co-ordinate system.  We define $\nu(t)$ to be the true anomaly, the phase measured from the periapsis direction to the current orbital position as seen by the ellipse's main focus. The argument of periapsis, relative to a fixed axis, is denoted $\gamma(t)$ so that the phase of the binary relative to a fixed axis is given by $\phi(t) = \nu(t) + \gamma(t)$. In many cases, it is simplest to describe the binary position with respect to the periapsis direction in terms of a uniformly increasing angle, the mean anomaly $l(t)$ which increases uniformly through one radial period (from periapsis to periapsis).  While these quantities are most naturally defined in the Newtonian limit of no radiation reaction, they extend in a natural way to the post-Newtonian framework \cite{Arun:2007sg, Moore:2016qxz}.  Formulae for the evolution of these parameters have been calculated to 3PN order \cite{Arun:2007rg, Arun:2007sg}, although here we focus primarily on the leading-order evolution as that motivates our later waveform decomposition.

\begin{figure}[t]%
    \includegraphics[width=0.48\textwidth, valign=t]{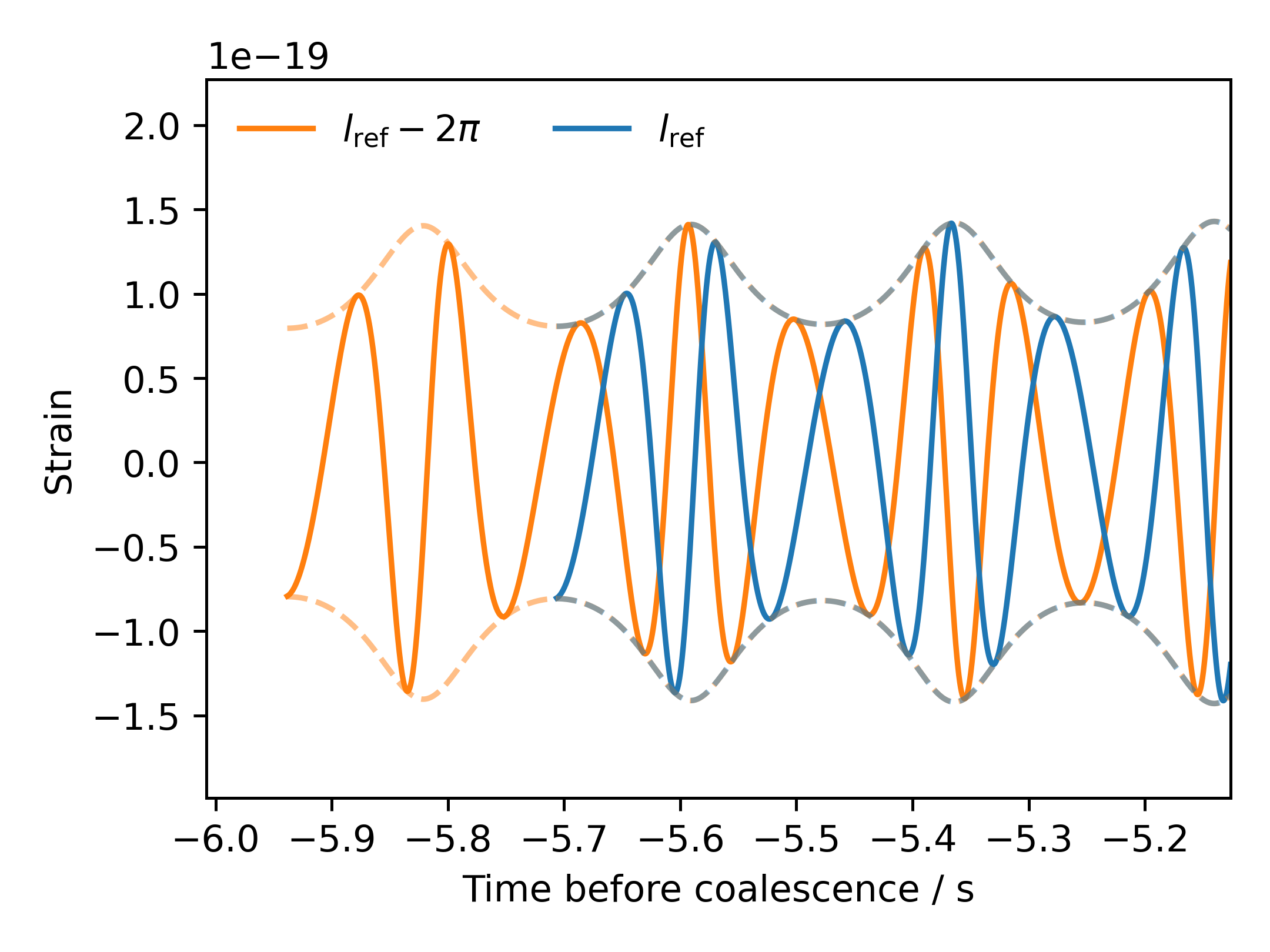}%
    \qquad
    \caption{A portion of the inspiral signal of a binary on an eccentric orbit with$f_\mathrm{ref}=\SI{10}{\Hz}, e_{10}=0.2$, $\mathcal{M}=\SI{24}{\Msolar}$, and $q=0.5$. The dashed envelope shows the overall amplitude of the waveform (including both $+$ and $\times$ \gls{gw} polarizations). 
    The orange curve shows an identical signal to the blue line but with the frequency and eccentricity back-evolved using Eq.~(\ref{Eq:shifted_e}) and Eq.~(\ref{Eq:shifted_f}) in order to start the waveform at the previous apoapsis.   The waveform was generated with the \texttt{TEOBResumS-Dali} \cite{Nagar:2021gss} model. 
    }%
    \label{Fig:EccWF}%
\end{figure}
  
At leading order, the gravitational wave emission occurs at twice the average orbital frequency $\omega = 2\pi f = 2 \left<\frac{d\phi} {dt} \right>$.  The evolution of the gravitational wave frequency with time is given as \cite{Peters:1963ux, Moore:2018kvz} (with $G=c=1$)

\begin{align}
\label{Eq:FreqEvol}
\left< \frac{df}{dt}\right> 
& = \frac{96\pi f^{2}}{5} \left(\pi \mathcal{M} f\right)^{5/3} 
\left(\frac{1+\frac{73}{24}e^2+\frac{37}{96}e^4}{\left(1-e^2\right)^{7/2}}\right) \, ,
\end{align}
and the evolution of the eccentricity as \cite{Peters:1963ux, Moore:2018kvz}
\begin{equation}\label{Eq:eccentricity_decay}
\left< \frac{de}{dt}\right> 
= -\frac{304}{15} \frac{1}{\mathcal{M}} \frac{\left(\pi \mathcal{M} f\right)^{8/3}}{(1 - e^{2})^{5/2}} 
\left(1 + \frac{121e^{2}}{304} \right) \, .
\end{equation}
These expressions can be combined to give the evolution of eccentricity with frequency, which can be integrated to \cite{Moore:2016qxz}
\begin{equation}
\label{Eq:shifted_e}
\frac{f}{f_{\mathrm{ref}}} = \left(\frac{e_{\mathrm{ref}}}{e}\right)^{18/19} \left(\frac{1-e^2}{1 - e_{\mathrm{ref}}^{2}}\right)^{3/2} \left(\frac{304 +121e_{\mathrm{ref}}^2}{304 +121 e^2}\right)^\frac{1305}{2299} \, .
\end{equation}
Thus, for moderate eccentricities, the second and third terms are close to unity and can be neglected and, to a reasonable approximation, the eccentricity decays as
\begin{equation}
    e \propto f^{-19/18} \, .
\end{equation}
Meanwhile the orbital separation reduces as $a \propto f^{-2/3}$, leading to circularization of the binary as frequency increases.

In Fig.~\ref{Fig:EccWF} we show a portion of the gravitational waveform emitted by an eccentric binary with with $\mathcal{M} =\SI{24}{\Msolar}$, mass ratio $q = 0.5$, beginning at a frequency $f_{\mathrm{ref}} = \SI{10}{\Hz}$ with an eccentricity $e_{\mathrm{ref}} = 0.2$, the argument of periapsis $\gamma = \pi$, and a mean of anomaly of $l = \pi$, so at the initial time the binary is at apoapsis and aligned with the fixed coordinate $x-$axis.  As expected, the amplitude of the waveform is modulated with the minimum signal at apoapsis and maximum at periapsis.  
In addition, the instantaneous
frequency of the emitted \glspl{gw} varies around the orbit with higher frequency when the amplitude is higher, as the black holes orbit faster around periapsis.  However, due to periapsis advance, the length of an amplitude modulation is noticeably longer than the time taken to complete one orbit, or equivalently two cycles of the \gls{gw} signal.  To demonstrate this more clearly, we also show the \gls{gw} signal from a binary which starts at apoapsis and is initially aligned with the $x-$axis one period earlier. The amplitude modulations in the two waveforms are aligned, but the phases of the waveform are offset due to periapsis advance.

The \gls{gw} phase oscillates twice over the course of the binary's azimuthal orbit (returning to a fixed orientation), while the amplitude modulations occur once per radial orbit (periapsis to periapsis).  Thus, we might expect to be able to decompose the signal into two distinct frequencies which we define as
\begin{equation}\label{Eq:freqs}
    \omega_\mathrm{r} = \frac{dl}{dt} 
    \quad \mbox{and} \quad
    \omega_{\phi} = \left< \frac{d \phi }{dt} \right> = 
    \omega_\mathrm{r} + 
    \left<\frac{d \gamma}{dt}\right> .
\end{equation}

The evolution of the periapsis is given at leading order for a non-spinning system by \cite{Einstein:1915bz, Moore:2016qxz}
\begin{equation}
\label{Eq:ApsidalAdvance}
\left< \frac{d \gamma}{dt}\right> = 
\frac{3 \omega_{\phi}^{5/3}\left(m_1+m_2\right)^{2/3}}{\left(1-e^2\right)} .
\end{equation}
Thus, the radial and azimuthal frequencies can be related through
\begin{equation}
\label{Eq:f_rad_f_az}
f_\mathrm{r} = 
f_\phi \left(1 - \frac{\Delta \gamma}{2\pi}\right) 
\quad \mbox{where} \quad
\Delta \gamma = \frac{1}{f_{\phi}} \left<\frac{d  \gamma}{dt}\right> \, .
\end{equation}
Here, $\Delta \gamma$ is the periapsis advance during one orbit.  By definition $f_{\phi}$ is always greater than $f_\mathrm{r}$ for eccentric systems, due to periapsis advance.  Higher order post-Newtonian expressions for these quantities are provided in, e.g. \cite{Moore:2016qxz}.

The leading-order waveform for an eccentric binary can be decomposed into a series of components each with a characteristic frequency determined by multiples of the radial and azimuthal frequencies \cite{Moreno:1995, Seto:2001pg}.  Observation of two or more of these components enables measurement of the orbital eccentricity, as has been discussed in the context of LISA observations \cite{Seto:2001pg, Willems:2007nq, Valsecchi_2012}.  The two polarizations of the emitted gravitational wave can be written at leading order, and assuming a uniform evolution of the periapsis, as%
\begin{align}
h_{+} &= - h_{o}(t) \frac{1 + \cos^{2} \iota}{2} 
\sum_{n} 
\left[ 
u_{n}(e) \cos(n l(t) + 2 \gamma(t)) \right. 
\nonumber \\
& \qquad \left.
+ v_{n}(e) \cos(n l(t) - 2 \gamma(t)) \right], 
\label{Eq:hp} \\ 
h_{\times} &= h_{o}(t) \cos \iota \sum_{n} 
\left[ 
u_{n}(e) \sin(n l(t) + 2 \gamma(t)) \right. \nonumber \\
& \qquad
+ \left. 
v_{n}(e) \sin(n l(t) - 2 \gamma(t)) \right].
\label{Eq:hc}
\end{align}
The functions $u_n$ and $v_n$ are calculated in \cite{Peters:1963ux, Moreno:1995} as combinations of Bessel functions and provided as power series in eccentricity in \cite{Seto:2001pg}. The only functions which are non-vanishing at $O(e^{0})$ or $O(e^{1})$ are for $n \le 3$ and these are given, up to and including $e^{2}$ terms, as
\begin{align}\label{Eq:ecc_amps}
    u_{1} &= - \frac{3e}{4} \, ,
    & 
    u_{2} &= 1 - \frac{5 e^2}{2} \, , 
    &
    u_{3} &= - \frac{9e}{4} \, , 
    \nonumber \\
    v_{1} &= 0 \, , 
    &
    v_{2} &= 0 \, , 
    &
    v_{3} &= 0 \,.
\end{align}
Thus, at zeroth order in eccentricity, the waveform is emitted at twice the azimuthal frequency, $f_{0} =  2 f_{\phi}$. The frequencies which appear at first order in eccentricity are $2f_\phi \pm f_\mathrm{r}$. 

The frequency structure of the waveform emitted by an eccentric binary can be understood at an intuitive level from the waveform in Fig.~\ref{Fig:EccWF}.  The primary gravitational wave emission occurs at twice the azimuthal frequency, as the gravitational wave strain is determined by (derivatives of) the quadrupole moments of the system.  Thus, the gravitational wave phase depends on the relative orientation of the two bodies.  The amplitude of the signal is largest at periapsis as the bodies are moving fastest and smallest at apoapsis when the bodies are moving more slowly.  Thus, the amplitude modulations depend upon the radial frequency.  It follows that the gravitational wave will have a leading component at twice the azimuthal frequency, with sub-leading contributions whose frequencies are increased/decreased by the radial frequency.  

The signal observed in a gravitational-wave detector depends upon the orientation of the binary relative to the detector.  Specifically, it depends upon the location and polarization, encoded in the detector response function, and binary inclination $\iota$.  The signal also depends upon the initial orientation of the binary.  This is parametrized by two angles, which specify the phase of the binary and the argument of periapsis.  The orientation can be specified by any two of the initial phase $\phi_{\mathrm{ref}}$, initial anomaly (either mean $l_{\mathrm{ref}}$ or true $\nu_{\mathrm{ref}}$) and argument of periapsis $\gamma_{\mathrm{ref}}$.  These are related to the initial true anomaly using $\phi_\mathrm{ref} = \nu_\mathrm{ref} + \gamma_\mathrm{ref}$, where there are well-known expressions to translate between mean and true anomaly.

In the above discussion, and for the remainder of the paper, we restrict attention to the (2, 2) multipole of the gravitational wave signal.  There is a contribution to the (2, 0) multipole at $O(e)$ with a frequency $f_{r}$\cite{Moreno:1995} which we will neglect. Its amplitude has a $\sin^{2} \iota$ dependence on inclination and will therefore vanish for face-on systems where the amplitude of the (2, 2) multipole is maximized.  The binary will emit in higher multipoles \cite{Ramos-Buades:2021adz, Nagar:2021gss, Islam:2024rhm, Islam:2024zqo}. We neglect these contributions as the most significant (3, 3) and (4, 4) modes vanish for face-on systems. Therefore, it is reasonable to expect that for most observed systems, which are preferentially viewed face-on, these contributions will be subdominant, however we plan to investigate their impact in a future publication.

\subsection{Eccentric harmonic frequencies}
\label{SubSec:HarmFreqs}

It is common \cite{Peters:1964zz, Moore:2018kvz} to decompose the eccentric binary waveform in multiples of the azimuthal frequency.  This basis has the advantage that the harmonics are naturally orthogonal.  However, the power, particularly at moderately high eccentricities, is spread over a large number of waveform harmonics.  As is clear from Fig.~\ref{Fig:EccWF} and the discussion above, the radial frequency naturally imparts a modulation onto the leading order waveform.  Therefore, it is also natural to decompose the waveform into a series
\begin{equation}\label{Eq:HarmFreqs}
    f_{k} = 2f_\phi + k f_\mathrm{r}\, ,
\end{equation}
as has been noted in \cite{Moreno:1995, Seto:2001pg, Valsecchi_2012, Moore:2016qxz}.

\begin{figure}[t]
\includegraphics[width=0.99 \linewidth]{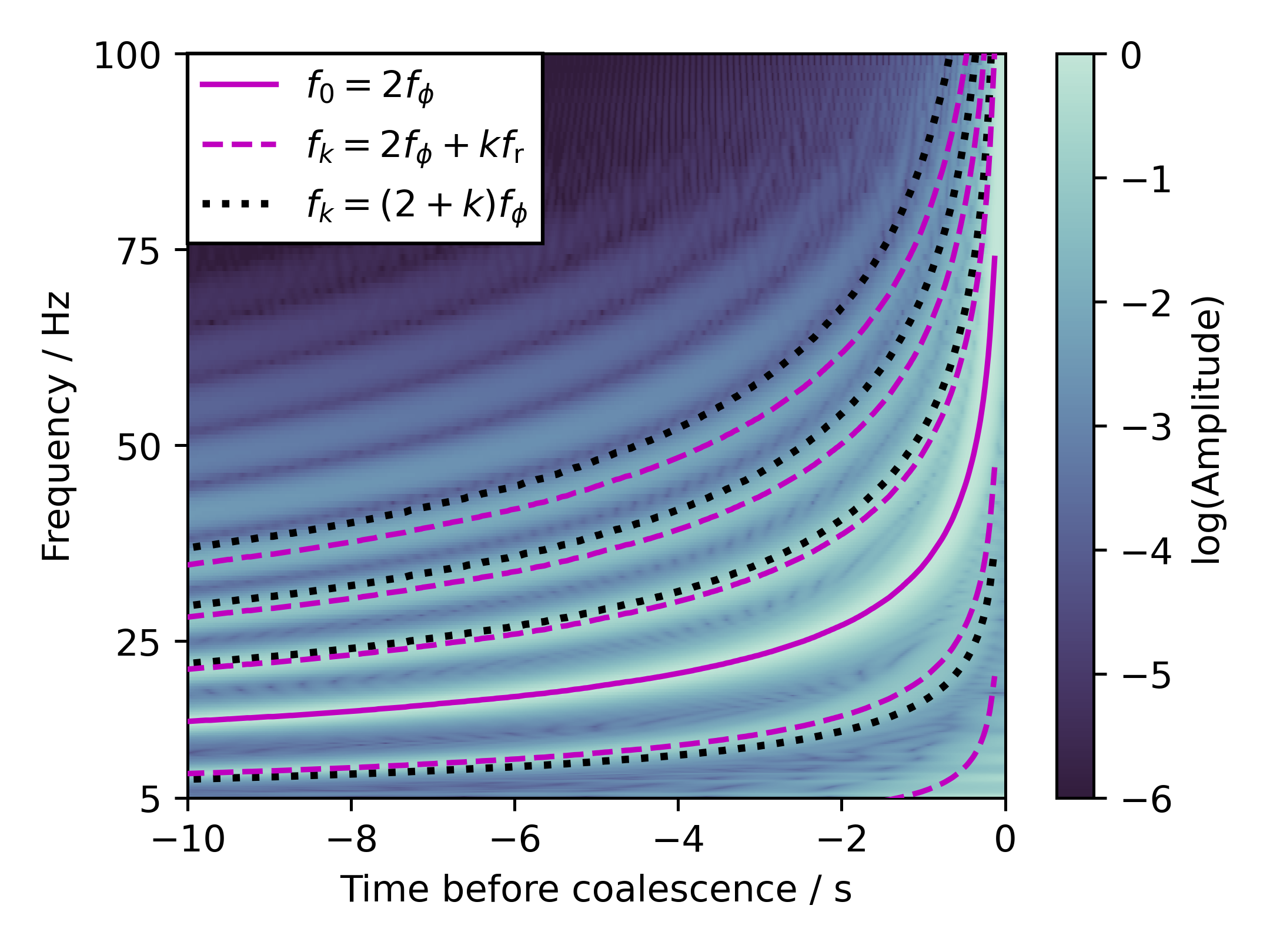}
\caption{Q-transform of a gravitational waveform from an eccentric binary, generated with the \texttt{TEOBResumS-Dali} model \cite{Nagar:2021gss}, with $f_\mathrm{ref}=\SI{5}{\Hz}, e_{10}=0.2$, $\mathcal{M}=\SI{10}{\Msolar}$, and $q=0.5$ with amplitude normalised over time by dividing all amplitudes by the maximum value at each frequency. The line of maximum amplitude at each time step is shown in purple, corresponding to the $h_0$ harmonic. The dashed purple and dashed black lines are higher harmonic frequency predictions from equation~(\ref{Eq:HarmFreqs}) and the integer multiple model respectively.}
\label{Fig:qtransform}
\end{figure}

Figure \ref{Fig:qtransform} shows a Q-transform of a signal from a black hole binary on an eccentric orbit.  
The frequency harmonics are clearly visible as curves of high amplitude whose frequencies increase as the binary approaches merger.  The loudest mode, marked with the solid line, has a frequency of twice the azimuthal frequency. The frequency of the other harmonics increases with time but their amplitudes, relative to the leading mode, decrease during inspiral as eccentricity is radiated away and the orbit circularises.

We also show lines of frequency $2f_{\phi} + k f_{r}$ and $(2 + k) f_{\phi}$ on the figure.  To do so, we first obtain the azimuthal frequency as $f_\phi = \tfrac{1}{2} f_0$.  We get the radial frequency by first calculating the eccentricity as a function of frequency from Eq.~(\ref{Eq:shifted_e}), using this to calculate the rate of of periapsis advance from Eq.~(\ref{Eq:ApsidalAdvance}) and the radial frequency from Eq.~(\ref{Eq:f_rad_f_az}).  The lowest frequency harmonic shown, $h_{-2}$, is at a very low frequency for the majority of the signal, however may enter the detector's sensitive band at late times.  Frequencies which are multiples of $f_{\phi}$ are shown as dotted lines on the figure, while those which vary by multiples of $f_{r}$ are shown with dashed lines.  It is clear the frequencies $2f_{\phi} + k f_{r}$ more accurately follow the power of the signal.  A decomposition in this frequency basis will then lead to power in a fewer modes than if we use multiples of $f_{\phi}$. 

The rate of apsidal advance is not constant over the merger of two compact bodies but increases as the frequency increases, as can be seen from Eq.~(\ref{Eq:ApsidalAdvance}).  Thus, the fractional difference between the two sets of frequencies increases as the system inspirals.  The inclusion of the apsidal advance is increasingly important to efficiently describe eccentric \gls{gw} signals close to merger, and allows for the existence of additional harmonic $h_{-2}$ (and even $h_{-3}$ at high frequencies).  As the binary approaches the \gls{isco}, $a_\mathrm{ISCO} = 6M$, we obtain, from Eq.~(\ref{Eq:ApsidalAdvance}) and Eq.~(\ref{Eq:f_rad_f_az}),
\begin{equation}
\label{Eq:ISCO_freqs}
f_\mathrm{r,ISCO} = f_{\phi,\mathrm{ISCO}} \left(1 - \frac{1}{2(1-e_\mathrm{ISCO}^2)} \right) 
\approx \frac{f_{\phi,\mathrm{ISCO}}}{2} \, ,
\end{equation}
where we have taken the eccentricity at \gls{isco} to be negligible which is a good approximation for most realistic choices of eccentricity.

\section{Waveform decomposition}
\label{Sec:Decomposition}

The waveform for an eccentric binary is composed of a set of harmonics at frequencies $2 f_{\phi} + k f_{\mathrm{r}}$.  However, waveform models for eccentric binaries \cite{Nagar:2024dzj, Ramos-Buades:2021adz} do not generate these harmonics, but rather the emitted waveform for a given set of parameters.  In this section, we present a method to generate a set component waveforms $h_{k}$, which contain the harmonic with frequency $f_{k} = 2f_\phi + k f_\mathrm{r}$.  To do so, we generate a set of eccentric waveforms, with identical masses, spins and eccentricity but varying initial phase and argument of periapsis.  By combining them appropriately, we obtain the harmonics $h_{k}$.  Then we can generate an eccentric waveform $h$ with an arbitrary phase and argument of periapsis as
\begin{equation}
\label{Eq:get_h}
h(\phi_\mathrm{ref}, \gamma_\mathrm{ref}) 
= \sum_k A_{k}(\phi_\mathrm{ref}, \gamma_\mathrm{ref}) h_{k},
\end{equation}
where $A_{k}$ are complex coefficients determining the contribution of each harmonic. 

\subsection{Generating the required waveforms}
\label{Sec:gen_wf}

To obtain the harmonics $h_{k}$, we wish to generate a set of waveforms $x_{j}$ with mean anomaly evenly spaced between $0$ and $2\pi$.  Unfortunately, this simple approach does not work as binaries with identical parameters other than the initial mean anomaly take different amounts of time to merge: a waveform starting close to periapsis, where the emitted \gls{gw} signal is maximal, will merge more quickly than one starting near apoapsis.  Therefore, we instead generate all waveforms with $l_{\mathrm{ref}}=\pi$, i.e. at apoapsis, and vary the initial frequency and eccentricity to ensure that the waveforms have the appropriate mean anomaly a fixed time before merger. 
This is a similar approach to the one introduced in \cite{Clarke:2022fma}.  We briefly describe the method below.

First, we calculate the change in frequency over on orbit from Eq.~(\ref{Eq:FreqEvol}), approximating the frequency and eccentricity as constant over a single orbit.  This gives the change in gravitational wave frequency in one azimuthal orbit as 
\begin{equation}
\label{Eq:FreqShiftOrbit}
\Delta f \approx
\frac{192 \pi f}{5} \left( \pi \mathcal{M} f\right)^{5/3} 
\left(\frac{1+\frac{73}{24}e^2+\frac{37}{96}e^4}{\left(1-e^2\right)^{7/2}} \right) \, .
\end{equation}
Due to periapsis advance, the binary must complete
\begin{equation}
\label{Eq:norbit}
n_\mathrm{orbit} = \left(1 - \frac{\Delta \gamma}{2\pi}\right)^{-1} 
\end{equation} 
orbits before returning to periapsis.  
Therefore, to generate waveforms which are evenly spaced in mean anomaly, we generate a set of waveforms all starting at apoapsis with initial frequencies
\begin{equation}
\label{Eq:shifted_f}
f_{j} = f_\mathrm{ref}
- \left(\frac{j}{n} \right) n_\mathrm{orbit}\Delta f 
,
\end{equation}
where $j = \{0,1,\ldots,n-1\}$.

While the binary evolves from $f_{j}$ to $f_{\mathrm{ref}}$, the eccentricity will decrease.  We use Eq.~(\ref{Eq:shifted_e}) to calculate the appropriate initial eccentricity to ensure an eccentricity $e_\mathrm{ref}$ at $f_\mathrm{ref}$.  While these expressions are derived at leading post-Newtonian order, we find they are sufficiently accurate to generate the required waveforms and extract the eccentric harmonics.

Figure~\ref{Fig:EccWF} shows waveforms which start exactly one radial orbit apart. The second waveform starts at the previous apoapsis and experiences one full amplitude modulation before the envelope joins up with the original waveform. The phase of the gravitational waveform differs for the two waveforms: both are generated with $\phi_\mathrm{ref} = 0$ but different initial frequencies and eccentricities. 

In principle, we can calculate the appropriate initial phase, $\phi_{j}$, for each waveform from the periapsis advance formula.  However, this is complicated by the fact that the phase does not evolve uniformly through the orbit --- it has both a secular and oscillatory contribution \cite{Arun:2007rg}.  Instead, we calculate the appropriate phase by directly comparing the gravitational waveforms.  To do so, we first combine both \gls{gw} polarizations into a single complex waveform using $x = x_+ - ix_\times$. We can then calculate the complex inner product $(x_{j} | x_{0})$ between the two waveforms, defined as
\begin{equation}
\label{Eq:InnerProd}
(a|b) = 4\int_{f_\mathrm{low}}^{f_\mathrm{high}}
\frac{\tilde{a}(f)\tilde{b}^\star(f)}{S(f)} df.
\end{equation}
Here $S(f)$ is the \gls{psd}%
\footnote{In this work we use the \gls{psd} corresponding to Advanced LIGO at design sensitivity with broad-band signal recycling and high power \cite{Abbott:2008} available at \href{https://dcc.ligo.org/ligo-t0900288/public}{dcc.ligo.org/ligo-t0900288/public}.}%
, and $\tilde{a}(f)$ denotes the Fourier transform of the signal $a(t)$.
The relative phase difference, $\phi_{j}$ between the two signals is obtained from $(x_{j}|x_{0}) = A \exp[2i \phi_{j}]$ and we can apply this phase offset to $x_{j}$ to obtain a signal which is in phase with $x_{0}$.

\subsection{Singular value decomposition}
\label{SubSec:SVD}

\Gls{svd} can used to identify the most important parameters describing a multivariate data set, reducing a large number of correlated basis vectors to a small number of orthogonal basis vectors which capture the key features of the data. \Gls{svd} has been previously applied to \gls{gw} data, for example to develop more efficient searches~\cite{Cannon:2010qh}. Here, we wish to use \gls{svd} to identify the key features of the eccentric waveforms and compare them to the theoretical predictions from Sec.~\ref{Sec:EccWaveform}.

A $n \times M$ complex matrix $\mathbf{X}$ is factorised as 
\begin{equation}
\label{Eq:SVD}
\mathbf{X} = \mathbf{U}^{\star}\mathbf{S}\mathbf{V},
\end{equation}
where $\mathbf{U}$ is a $n \times n$ complex unitary matrix, $\mathbf{S}$ is a $n \times M$ rectangular diagonal matrix of positive real numbers, and $\mathbf{V}$ is an $M \times M$ complex unitary matrix. 
Here, our matrix $\mathbf{X}$ is formed from $n=100$ gravitational wave signals from an eccentric binary system, $\{x_{0}, x_{1}, \ldots, x_{99}\}$, with identical masses and spins but varying mean anomaly, generated as described in Sec.~\ref{Sec:gen_wf}.
By construction, the length of all of the waveforms is equal and consequently M represents the number of time samples in each waveform. We first whiten these waveforms using the appropriate detector \gls{psd}, $S(f)$, to ensure that the \gls{svd} decomposition captures the most significant \textit{observable} features of the waveform. To do this, we Fourier transform the waveforms to the frequency domain, whiten by dividing by $\sqrt{S(f)}$, before returning to the time domain and performing \gls{svd}. 

The matrix $\mathbf{V}$ contains a set of $M$ waveform components, $h_{k}^\mathrm{SVD}$, which span the space of possible vectors of length $M$. Of these, the first $n$ provide basis waveforms which can be used to reconstruct the initial eccentric binary waveforms $x_{j}$, while the remaining rows are orthogonal to the first $n$ and ensure that $\mathbf{V}$ comprises a complete basis. In what follows, we only consider the first $n$ rows of $\mathbf{V}$.  These $h_{k}^{\mathrm{SVD}}$ are ordered in terms of importance, so that $h_{0}^{\mathrm{SVD}}$ is normalized waveform that describes the primary features of the original $\{x_{j}\}$.
The $\mathbf{S}$ matrix contains $n$ diagonal values, $S_{k}$, describing the relative importance of the corresponding waveform component $h_{k}^\mathrm{SVD}$.  In particular, the fraction of information contained in the waveform $h_{k}^{\mathrm{SVD}}$ is given by
\begin{equation}
\label{Eq:RelImport}
\alpha_{k} = 
\frac{S_{k}^2}{\sum_{k^{\prime}} S_{k^{\prime}}^2} \, .
\end{equation}
The matrix $\mathbf{U}^{\star}$ provides the weighted contribution of each of the \gls{svd} waveforms, $h_{k}^{\mathrm{SVD}}$, to the original set of waveforms $x_{j}$.  Equivalently, we can rewrite Eq.~(\ref{Eq:SVD}) as 
\begin{equation}
\label{Eq:SVD_V}
\mathbf{S}\mathbf{V} =  \mathbf{U} \mathbf{X} \, .
\end{equation}
Thus, the matrix $\mathbf{U}$ also gives the contribution of each of the $x_{j}$ to the basis waveforms $h_{k}^{\mathrm{SVD}}$. 

\begin{figure*}[ht]
\centering
\includegraphics[width=\linewidth, valign=b]{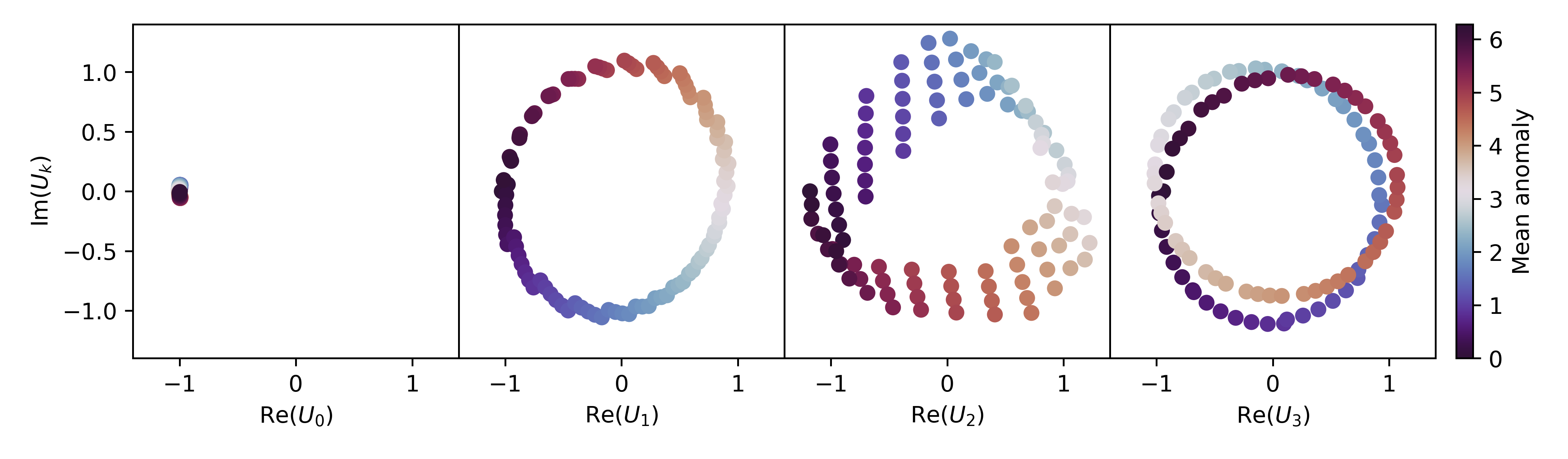}
\caption{Coefficients (arbitrary normalization) of component waveform used by the \gls{svd} to construct the component $h_{k}^{\mathrm{SVD}}$ waveform space for $e_{10}=0.2$, $\mathcal{M}=\SI{24}{\Msolar}$, and $q=0.5$. The colours depict initial mean anomaly of the waveform at $t_\mathrm{ref}$, which varies from $0$ to $2\pi$. The four components here are those containing the most information, sorted in descending order from left to right. The most important component has an equal contribution from each waveform, meaning it is simply the average.  The other components use an (essentially) equal magnitude contribution from each component but a phase which varies by $2\pi$, $-2\pi$ and $4\pi$ respectively for the second, third, and fourth most important components.
}
\label{Fig:SVD100_coeffs}
\end{figure*}

In generating the \gls{svd} component waveforms, we have used \textit{whitened} waveforms $x_{j}$.  However, in many cases it is useful to construct \textit{un-whitened} waveforms associated with the \gls{svd} basis.  We can easily do this using Eq.~(\ref{Eq:SVD_V}).  Given the coefficients $\mathbf{U}$ and normalizations $\mathbf{S}$ obtained using whitened waveforms, we can use unwhitened waveforms to construct a new matrix $\mathbf{X}$ and generate the un-whitened \gls{svd} basis $\mathbf{V}$.

In Fig.~\ref{Fig:SVD100_coeffs} we show the coefficients for the contribution of each $x_{j}$ to the first four \gls{svd} components. The zeroth \gls{svd} waveform has an equal contribution from each waveform, so that it averages over the modulations caused by eccentricity.  The next three components have an approximately equal magnitude contribution from each waveform but a phase which varies by $2\pi$, $-2\pi$ and $4\pi$ respectively. Each factor of $2\pi$ in phase of the \gls{svd} coefficients corresponds to an increase (or decrease) of the waveform frequency by $f_{\mathrm{r}}$. Thus we recognise these components as corresponding to the $k = 0, 1, -1, 2$ gravitational wave harmonics introduced in Sec.~\ref{Sec:EccWaveform}. A subset of higher \gls{svd} components can also be mapped to the eccentric harmonics while others cannot be clearly identified --- likely because there is not significant power in more than the leading few harmonics.  

We can use Eq.~(\ref{Eq:RelImport}) to assess the relative importance of the \gls{svd} components.
For a signal with $e_{10}=0.2$ and $\mathcal{M}=\SI{24}{\Msolar}$, we find that 98.6\% of the total information is contained in just the first two components, and 99.9\% contained in the first four. This demonstrates that we require a small number of components to accurately describe eccentric waveforms.  We investigate this in more detail in Sec.~\ref{SubSec:HarmonicPlots} after first introducing a faster way to generate the eccentric harmonics.

\subsection{Constructing eccentric harmonics}
\label{SubSec:EccHarms}

Only a small number of the \gls{svd} components are needed to accurately describe the gravitational waveform emitted by an eccentric binary with fixed masses and spins but arbitrary initial orientation, as determined by the argument of periapsis and true anomaly.  Generating this basis using an \gls{svd} decomposition of 100 eccentric waveforms is unnecessarily complex and time-consuming.  Given the frequency structure of the eccentric harmonics predicted theoretically in Eq.~(\ref{Eq:HarmFreqs}) and confirmed in the \gls{svd} decomposition, we propose a more straightforward and computationally efficient method to generate them.

From Fig.~\ref{Fig:SVD100_coeffs}, it follows that the eccentric harmonics can be generated by summing the waveforms $x_{j}$ with the appropriate phase factor as
\begin{equation}\label{Eq:hk}
h_{k} = \frac{1}{n} 
\sum_{j=0}^{n-1} e^{(2 \pi i jk/n)} x_{j} \, .
\end{equation}
As before $x_{j}$ are waveforms generated evenly spaced in mean anomaly at a fixed time before merger, $n$ gives the total number of waveforms and $k$ is the index labelling the desired harmonic.
This ensures that the frequency of $h_{k}$ is given by $f_{k} = 2 f_\phi + k f_{\mathrm{r}}$.  Since we are interested in only the leading few harmonics, we can use a small value of $n$ and still recover accurate representations of the waveform.  It is easy to see that, by definition, $h_{k} = h_{k\pm n}$. Therefore if $n$ is larger than the number of eccentric harmonics with significant power, each $h_{k}$ will only contain power from one harmonic.
Unless otherwise specified we choose to use $n = 6$ in this work as there is limited power in the $k=-2$ harmonic owing to its very low frequency content, and the $k=5$ harmonic has negligible power over the range of masses and eccentricities used in this paper. Therefore, we generate eccentric harmonics corresponding to $k= -1, 0, 1, 2, 3,$ and $4$.

The waveforms generated by the \gls{svd} are, by definition, orthogonal, but this is not the case for the harmonics obtained using Eq.~(\ref{Eq:hk}). When representing a waveform as a sum of appropriately weighted harmonics in Eq.~(\ref{Eq:get_h}), it is simpler to calculate the coefficients if the waveform components are orthogonal. We therefore apply Gram-Schmidt orthogonalization to obtain a set of orthogonal waveforms. We iteratively orthogonalize the waveforms by defining
$h_l^\perp$ as
\begin{equation}
\label{Eq:GSOrthogonalised}
h^\perp_{l} = h_{l} - \sum_{i =0}^{l-1} \frac{(h_{l}|h_{i})}{(h_{i}|h_{i})} h_{i} \, ,
\end{equation}
where the index $l$ relabels the harmonics in order of descending importance, i.e. harmonics with $l = \{0, 1, 2, 3, 4, \ldots\}$ correspond to harmonics with $k = \{0, 1, -1, 2, 3, \ldots\}$, such that the sum runs over the indices of only more important harmonics. From now on we will drop the $^\perp$ notation for convenience. 
For most of the parameter space the orthogonalization has little impact on the waveforms, however for high mass signals the observable waveform is short and the binary completes only a few orbits in band.  Therefore, the different harmonics can have large overlaps and the projected waveforms can differ significantly from the original ones. In this limit, it becomes difficult to robustly identify the individual harmonics.

\begin{table}[t]
\caption{Overlap between our harmonic waveforms ($h_{k}$) generated with $n=\numrange{2}{6}$ waveforms ($x_{k}$) and waveforms constructed from a \gls{svd} analysis performed on 100 component waveforms ($h_{k}^\mathrm{SVD}$) with $e_{10}=0.2$, $\mathcal{M}=\SI{24}{\Msolar}$, and $q=0.5$.
}
\centering
\begin{tabular}{cccccc}
\hline
$k$ & \multicolumn{1}{r}{$n=2$} & $n=3$  & $n=4$  & $n=5$ & $n=6$  \\ \hline
0  & 0.9963  & 0.9996 & 0.9991 & 0.9996 & 1.0000 \\
1  & 0.9225  & 0.9949 & 0.9995 & 0.9993 & 0.9995 \\
-1 & -       & 0.8050 & 0.9618 & 0.9917 & 0.9908 \\
2  & -       & -      & 0.9946 & 0.9957 & 0.9621 \\
3  & -       & -      & -      & 0.9667 & 0.9700 \\
4  & -       & -      & -      & -      & 0.8399
\end{tabular}
\label{Tbl:SV100_match}
\end{table}

In Table~\ref{Tbl:SV100_match} we show the overlap between the \gls{svd} waveforms and those generated using Eq.~(\ref{Eq:hk}), for different values of $n$.  The waveform overlap is defined as
\begin{equation}
\label{Eq:Overlap}
\mathcal{O}(h, h^\prime) = \frac{|(h|h^\prime)|}{|h||h^\prime|},
\end{equation}
where 
\begin{equation}
\label{Eq:NormOverlap}
|h| = \sqrt{(h|h)}, 
\end{equation}
so that $\mathcal{O}(h, h^\prime) = 1$ indicates identical waveforms, up to an overall phase. The harmonics are in good agreement with the \gls{svd} components in most cases.  There is a poor agreement for the $k=1$ component when $n=2$ and the $k=-1$ harmonic when $n=3$, likely due to the fact that two harmonics with observable power mapped to the same component. The $k=4$ waveform has limited power, which likely explains the poor agreement with the \gls{svd} component.  Other than these cases, the overlaps are greater than $0.95$ in all cases and $0.99$ for most.

\subsection{Generating eccentric harmonics}
\label{SubSec:HarmonicPlots}

\begin{figure*}[t]
\centering
\includegraphics[width=\linewidth, valign=b]{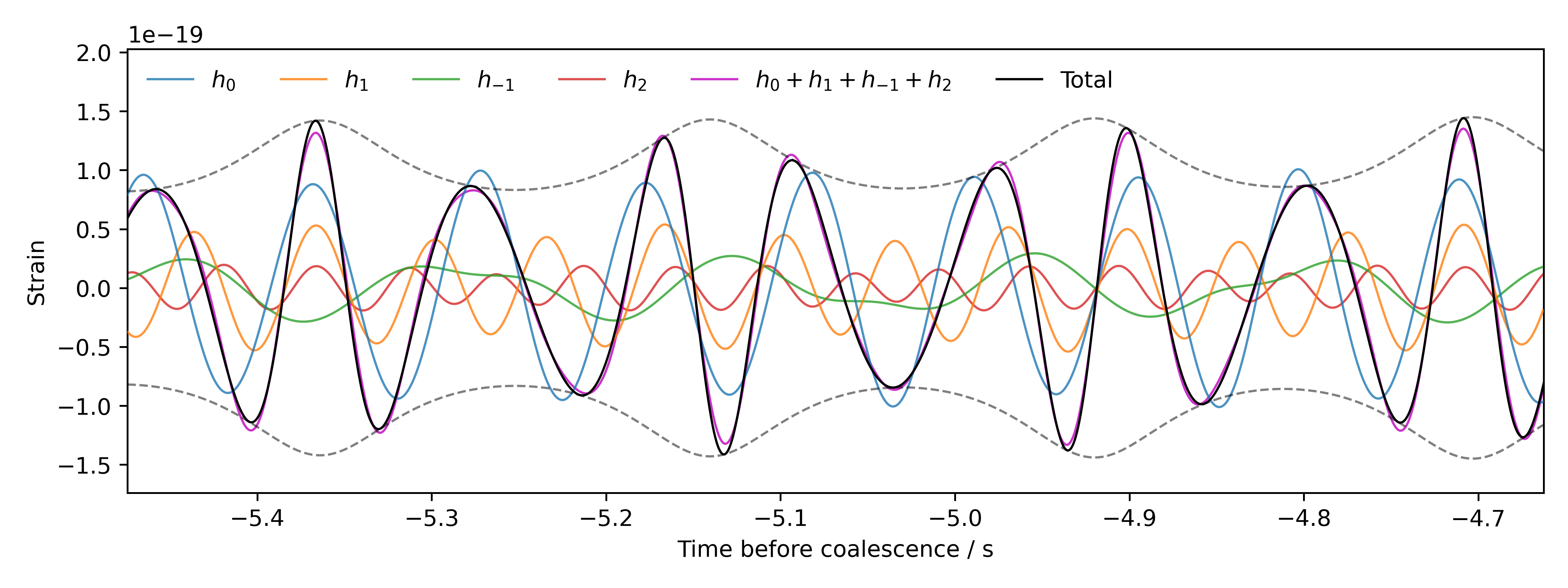}
\caption{The gravitational waveform (black) and the four leading harmonics for a signal with $e_{10}=0.2$, $\mathcal{M}=\SI{24}{\Msolar}$, and $q=0.5$ and vanishing component spins, starting at $f_\mathrm{ref}=\SI{10}{\Hz}$, generated using the \texttt{TEOBResumS-Dali} model \cite{Nagar:2021gss}. The harmonics are generated using Eq.~(\ref{Eq:hk}) with $n=10$. The magenta line shows the sum of these four harmonics, which is a good approximation to the full signal. 
}
\label{Fig:ecc_components}
\end{figure*}

Figure \ref{Fig:ecc_components} shows a portion of an eccentric waveform for $e_{10}=0.2$, $\mathcal{M}=\SI{24}{\Msolar}$, $q=0.5$. In addition to the full waveform, we show the contributions to the waveform from the four leading eccentric harmonics, $h_0$, $h_1$, $h_{-1}$ and $h_{2}$, constructed as described in Sec.~\ref{SubSec:EccHarms}.  We also show the approximate waveform generated from these harmonics, which is an excellent fit to the full waveform. This confirms both that a small number of harmonics can be used to accurately reconstruct the eccentric waveform and that our efficient method of generating these harmonics is sufficiently accurate.

\begin{figure*}[t]
\centering
\includegraphics[width=\linewidth, valign=b]{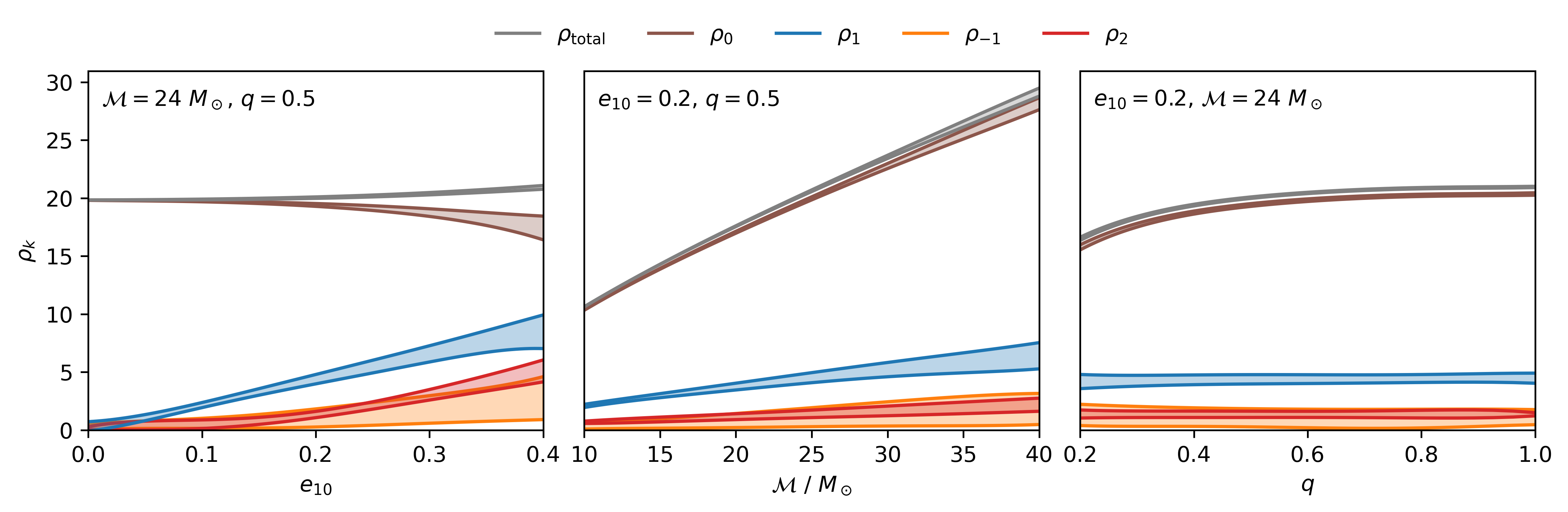}
\caption{The expected \gls{snr} for an eccentric binary merger.  The fiducial values of the eccentricity at $\SI{10}{\Hz}$, $e_{10}$, chirp mass,$\mathcal{M}$, mass ratio, q = $m_2/m_1$ are $e_{10}=0.2$, $\mathcal{M}=\SI{24}{\Msolar}$ and $q=0.5$.  The components are non-spinning and the binary is placed at a distance of $\SI{1680}{\mega \parsec}$ directly above a detector operating the Advanced LIGO design sensitivity. In each plot, we vary one parameter, \textit{left:} the eccentricity, \textit{centre:} the chirp mass and \textit{right:} the mass ratio, keeping other values fixed. 
The shaded regions denote the range of \glspl{snr} as the orientation (initial phase and argument of periapsis) is varied.  The grey region shows the total \gls{snr} in the signal while the  \{brown, blue, orange, red\} regions show the SNR of the k=\{0,1,-1,2\} harmonics respectively.
}
\label{Fig:harm_power}
\end{figure*}

In Fig.~\ref{Fig:harm_power}, we show how the \gls{snr} in the eccentric harmonics varies with eccentricity, chirp mass, and mass ratio. To do so, we generate the harmonics for a system with the specified chirp mass, mass ratio and eccentricity using Eq.~(\ref{Eq:hk}). We note that at fixed values of masses, spins, and eccentricity, the \gls{snr} in an eccentric merger (and in each of the harmonics) can vary with the initial orientation of the binary. Consequently, we generate multiple signals for each point in parameter space, each consisting of a waveform generated with a different initial mean anomaly. We therefore show the \glspl{snr} in Fig.~\ref{Fig:harm_power} as bands containing the range of possible \glspl{snr}. We calculate the \gls{snr} for each signal, $h$, in data, $d$, using~\cite{Allen:2005fk, Brown:2004vh, LIGOScientific:2016vbw}
\begin{equation}
\label{Eq:SNRInnerProd}
\rho = \mathrm{Max}_{t_{c}} \frac{|(h(t_c)|d)|}{|h|}\, ,
\end{equation}
where the complex matched filter is defined in Eq.~(\ref{Eq:InnerProd}). In all cases in this work we generate waveforms beginning at $\SI{10}{\Hz}$ but begin the matched filtering at $\SI{20}{\Hz}$. When matched filtering individual eccentric harmonics, we want to use the same value of $t_c$ for each harmonic to ensure they correspond to a binary with the same coalescence time. To achieve this we first calculate the \gls{snr} $\rho_0$ for the leading $k=0$ harmonic, and require that the identical time $t_c$ is used to calculate \glspl{snr} $\rho_k$. As there is no noise contribution in this case, $d=h$ and so $\rho_\mathrm{total} = |h|$. 

Figure \ref{Fig:harm_power} shows that the total \gls{snr} of a binary increases slightly with eccentricity --- although the binary will merge more quickly, the emitted waveform amplitude will be higher. The fraction of the \gls{snr} captured by the $k=0$ harmonic decreases as $e_{10}$ increases, with close to 100\% of the power contained in the leading harmonic at $e_{10} = 0$, reducing to $80\%$ at $e_{10} \approx 0.4$.  The majority of the additional power is captured by the $k=1$ harmonic, with increasing power in both the $k=-1$ and $k=2$ harmonics at higher eccentricities.  As expected from Eq.~(\ref{Eq:ecc_amps}), the \gls{snr} in the $k=1$ and $-1$ harmonics increases linearly with eccentricity while the falloff of \gls{snr} in $k=0$ and the growth of \gls{snr} in the $k=2$ harmonic vary quadratically.   

The \gls{snr} in the signal increases with increasing chirp mass, as the overall amplitude is higher.  However, there is little change in the relative importance of the harmonics, with the $k=0$ harmonic capturing the vast majority of the \gls{snr} over all masses and the $k=1$ harmonic being the second most significant.  As the mass ratio is varied, the total \gls{snr} of the signal reduces for $q \lesssim 0.5$, although the power in the $k=-1, 1$ and $2$ harmonics remains approximately constant, indicating that these harmonics become more important for unequal mass binaries.

\section{Degeneracy between eccentricity and mass}

\label{Sec:Degeneracy}

We have introduced a decomposition of the waveform emitted by an eccentric binary into a series of harmonics. Furthermore we have shown that the vast majority of the signal power is contained within the first few harmonics. In addition, as can be seen from Fig.~\ref{Fig:harm_power}, the significance of the $k\neq0$ harmonics increases with increasing eccentricity.  As expected from theoretical considerations, this increase is approximately linear for the $k=1$ and $-1$ harmonics.  This suggests that it may be possible measure the eccentricity using the relative \gls{snr} in the different harmonics.  For this approach to be viable, and computationally feasible, we require the eccentric harmonics to provide an accurate representation of the waveform over a range of eccentricities (in much the same way as a small number of precession harmonics can describe a precessing waveform with different values of in-plane spins \cite{Fairhurst:2019vut, McIsaac:2023ijd}).  It is well known, see e.g.~\cite{Favata:2021vhw}, that there exist degeneracies between the eccentricity and other parameters, most notably the chirp mass.  Here, we investigate whether the eccentric harmonics are, at least approximately, valid over a range of eccentricities if we appropriately change the values of the other parameters, specifically the chirp mass.

\subsection{Degeneracy of $h_{0}$ with chirp mass}
\label{Sec:deg_chirp}

\begin{figure}[t]
\centering
\includegraphics[width=\linewidth, valign=b]{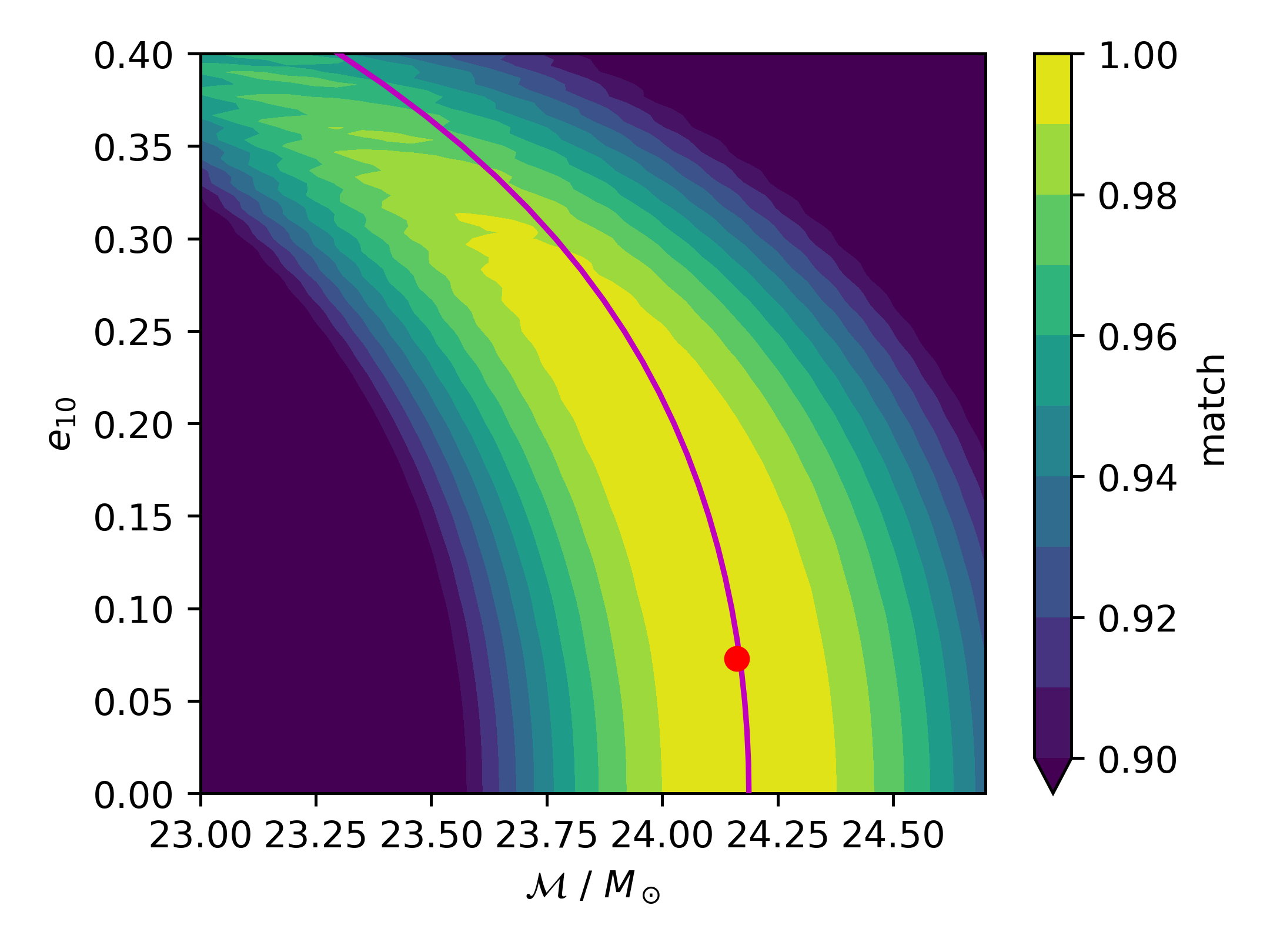}
\caption{Match between the $h_0$ eccentric waveform harmonic across the chirp mass and eccentricity space, calculated between the fiducial waveform, generated at $e_{10}=0.073, \mathcal{M}=\SI{24.14}{\Msolar}$ (indicated by the red dot and chosen such that our illustrative event lies on the degeneracy line) and the point shown. In all cases the system has a mass ratio of $q=0.5$, the waveform is generated from $\SI{10}{\hertz}$ and the match performed using the aLIGO \gls{psd} with a low frequency cutoff of $\SI{20}{\hertz}$. The magenta line shows the line of degeneracy between eccentricity and chirp mass described by Eq.~(\ref{Eq:DegeneracyEqn}).
}
\label{Fig:match_e_mchirp}
\end{figure}

In Fig.~\ref{Fig:match_e_mchirp}, we show how well the $h_{0}$ harmonic from a fixed signal matches with the equivalent harmonic for a second waveform with varying eccentricity and chirp mass. Our initial binary has an eccentricity $e_{10}=0.073$, $\mathcal{M}=\SI{24.14}{\Msolar}$, $q=0.5$, and both black holes have zero spin.  The match, $M$, is the maximum value of the overlap, maximized over time shifts as 
\begin{equation}
    M(h, h^{\prime}) = \mathrm{Max}_{\mathrm{\delta t}} \mathcal{O}(h(t_{c}), h^{\prime}(t_c + \delta t)) \, , 
\end{equation}
where $t_c$ and $t_c + \delta t$ denote the coalescence times of the two waveforms.  The match encodes the fraction of the \gls{snr} in the signal $h$ which is recovered when filtering with the waveform $h^{\prime}$.   To generate the eccentric harmonics, we use the procedure discussed in Sec.~\ref{SubSec:EccHarms} with $n=6$ waveforms. As discussed in Sec.~\ref{SubSec:HarmonicPlots}, we generate waveforms beginning at $\SI{10}{\Hz}$ and calculate the match between waveforms starting at $\SI{20}{\hertz}$. 

Figure \ref{Fig:match_e_mchirp} shows a clear degeneracy between eccentricity and chirp mass as expected.  Binaries with lower mass and higher eccentricity are largely indistinguishable from the initial signal. In \cite{Favata:2021vhw}, the authors introduced an eccentric chirp mass $\mathcal{M}_\mathrm{ecc}$ to map the lines of degeneracy and we follow that approach here.  To do so we work at leading post-Newtonian order and consider only the inspiral part of the waveform. In that case we can write the waveform as
\begin{equation}
h(f) = \mathcal{A}(f) e^{i\phi(f)},    
\end{equation}
where $\mathcal{A}(f) = \mathcal{A} f^{-7/6}$ and $\mathcal{A}$ 
depends upon the mass of and distance to the binary. We choose to neglect the dependence of the amplitude $\mathcal{A}(f)$ with eccentricity, as this will be sub-leading in comparison to the phasing \cite{Favata:2021vhw}. The phase can be written as a function of frequency as
\begin{equation}
\label{Eq:gw_phase}
\phi(f) = \phi_{c} + 2\pi f t_{c} +  a \mathcal{M}_\mathrm{ecc}^{-5/3}f^{-5/3},
\end{equation}
where $\phi_c$ and $t_c$ are the phase and time of coalescence respectively and $a = 3/ (128 \pi^{5/3})$.  The eccentric chirp mass $\mathcal{M}_{\mathrm{ecc}}$ depends upon the eccentricity and is defined as \cite{Favata:2021vhw}
\begin{equation}\label{Eq:mchirp_ecc}
\mathcal{M}_\mathrm{ecc}^{-5/3} = \mathcal{M}^{-5/3} \left(1-\frac{2355}{1462} e_f^2\right) 
=: \mathcal{M}^{-5/3} \left(1 - k e_f^2 \right) \, ,
\end{equation}
where $e_{f}$ is the eccentricity as a function of frequency $f$ and we have defined the constant $k = \tfrac{2355}{1462}$.

Since the eccentricity decreases with frequency, the eccentric chirp mass changes as the binary evolves.  To obtain a single value of $\mathcal{M}_{\mathrm{ecc}}$ we must choose a reference frequency at which to evaluate the eccentricity. In \cite{Favata:2021vhw}, the authors expand the phasing in powers of $(f - f_\mathrm{ref})$ to obtain an eccentric chirp mass which provides the dominant impact on the phasing at $f_\mathrm{ref}$. Here, we instead calculate a \textit{frequency averaged} eccentric chirp mass, where the frequency averaging is based upon the relative amplitude of the signal, $h(f)$, to the noise, $\sqrt{S(f)}$.  

We begin by evaluating the overlap between a non-eccentric signal with chirp mass $\hat{\mathcal{M}}$ and an eccentric signal with chirp mass $\mathcal{M}$ and initial eccentricity $e_\mathrm{ref}$.  By maximizing the overlap as a function of mass, we obtain the degeneracy between mass and eccentricity that appropriately weights varying effective chirp mass over the inspiral. The overlap between the two signals is%
\footnote{We fix the relative timing between the signals here by calculating an overlap rather than a match.  The result provides good agreement with the observed degeneracy, indicating that the maximization over time would have little impact on the result.}
\begin{equation}
\mathcal{O} = \left| \int df \frac{\mathcal{A}(f)^2}{S(f)} \exp
\left\{ iaf^{-5/3}\left(\hat{\mathcal{M}}^{-5/3} - \mathcal{M}_\mathrm{ecc}^{-5/3}\right)\right\} \right| \, .
\end{equation}
Next, we assume that the chirp mass $\mathcal{M}$ is close to the chirp mass, $\hat{\mathcal{M}}$, of the non-eccentric signal and expand at leading order as
\begin{equation}
\mathcal{M}_\mathrm{ecc}^{-5/3} 
= \hat{\mathcal{M}}^{-5/3}\left(1 + \delta m\right) \left(1 - k e_f^2\right) \, .
\end{equation}
This allows us to express the overlap as
\begin{equation}\label{Eq:over_dm}
\mathcal{O} \approx \left| \int df \frac{\mathcal{A}(f)^{2}}{S(f)} 
\exp\{ia(\hat{\mathcal{M}}f)^{-5/3}\left(\delta m - k e_f^2 ( 1 +\delta m) \right)\} \right|.
\end{equation}
We then expand the exponential including terms up to quadratic order in either $\delta m$ or $e_f^2$.
To do so, we square Eq.~(\ref{Eq:over_dm}) and expand, obtaining a quadratic contribution from either the product of first-order terms or a single second order term.  Thus, we obtain
\begin{align}
\mathcal{O}^2 \approx 1 - a^2 \hat{\mathcal{M}}^{-10/3} & \left[ \,
\overline{f^{-10/3} \left(\delta m - k e_f^2 \right)^2} 
\right.
\nonumber \\
&\left.
-
\left( \overline{f^{-5/3}\left(\delta m - k e_f^2\right)} \right)^2  
\right]
\, ,
\end{align}
where we have defined $\overline{x}$ as the frequency averaged value of $x$, 
\begin{equation}
\left.
\overline{x} =  4 \int df \frac{x\, |h|^2}{S(f)}  
\middle/
4 \int df \frac{|h|^2}{S(f)} 
\right.
\, .
\end{equation}
Finally, we collect terms with equal powers of $\delta m$ to obtain
\begin{align}\label{Eq:mismatch}
1 - \mathcal{O}^2 &\approx a^2 \hat{\mathcal{M}}^{-10/3} \left\{
\delta m^2 \left[\overline{f^{-10/3}} - \left(\overline{f^{-5/3}}\right)^2 \right] \right. \nonumber\\ 
& \quad - 2 \delta m \left[\overline{ ke_{f}^{2}f^{-10/3}} 
- \left(\overline{ke_{f}^{2}f^{-5/3}}\right) \left(\overline{f^{-5/3}} \right)
\; \right] \nonumber\\
& \quad \left. + \left[ \overline{k^{2}e_{f}^{4} f^{-10/3}} 
- \left(\overline{ke_{f}^{2}{f}^{-5/3}}\right)^2 \right]\right\}.
\end{align}
In order to find the degeneracy, we differentiate with respect to $\delta m$ to obtain
\begin{equation}\label{Eq:delta_m}
\delta m = \frac{
\overline{ ke_{f}^{2} f^{-10/3}} 
- \left(\overline{ ke_{f}^{2} f^{-5/3}}\right) \left(\overline{f^{-5/3}}\right)  
}{\overline{f^{-10/3}} - \left(\overline{f^{-5/3}}\right)^{2}}.
\end{equation}
Substituting the expression for eccentricity as a function of frequency, Eq.~(\ref{Eq:shifted_e}), gives the degeneracy between $\delta m$ and reference eccentricity, $e_{10}$.   Finally, we obtain the chirp mass $\mathcal{M}$ as a function of eccentricity as
\begin{equation}
\label{Eq:DegeneracyEqn}
\mathcal{M} = \hat{\mathcal{M}} \left(1+\delta m\right)^{-3/5},
\end{equation}
where $\delta m$ is given in Eq.~(\ref{Eq:delta_m}).  The degeneracy line is plotted on Fig.~\ref{Fig:match_e_mchirp} and is in excellent agreement with the observed degeneracy.

For a signal observed at \gls{snr} around 10, the 90\% confidence interval for parameter recovery approximately corresponds to the region with match $\ge 0.97$ \cite{Baird:2012cu}.  This extends to an eccentricity of close to $0.4$.  Therefore, for a low \gls{snr} observation, the leading order eccentric harmonic is appropriate for relatively large eccentricities.

It is also useful to understand how the extent of the $\mathcal{M}$ and $e_{10}$ degeneracy will scale with mass.  To do so, we must identifying how a contour of fixed overlap scales with the chirp mass. It is clear that substituting $\delta m$ from Eq.~(\ref{Eq:delta_m}) into Eq.~(\ref{Eq:mismatch}) gives a series of terms which all depend upon the combination of $\mathcal{M}^{-10/3} e_f^4$ (although the specific frequency weighting of each term does vary). Then, using the leading order relation between eccentricity and \gls{gw} frequency (the first term in Eq.~(\ref{Eq:shifted_e})) we see that eccentricity is (approximately) proportional to frequency. This allows us to conclude that for low eccentricities and relatively low masses, where the inspiral part of the waveform is the dominant contribution, that a contour of fixed overlap requires a fixed value of $\hat{\mathcal{M}}^{-10/3} e_\mathrm{10}^4$. Equivalently, we expect results to scale as $e_\mathrm{10} \propto \hat{\mathcal{M}}^{5/6}$.

\subsection{Degeneracy for other eccentric harmonics}
\label{Sec:deg_harms}

\begin{figure*}[t]
\centering
\includegraphics[width=\linewidth, valign=b]{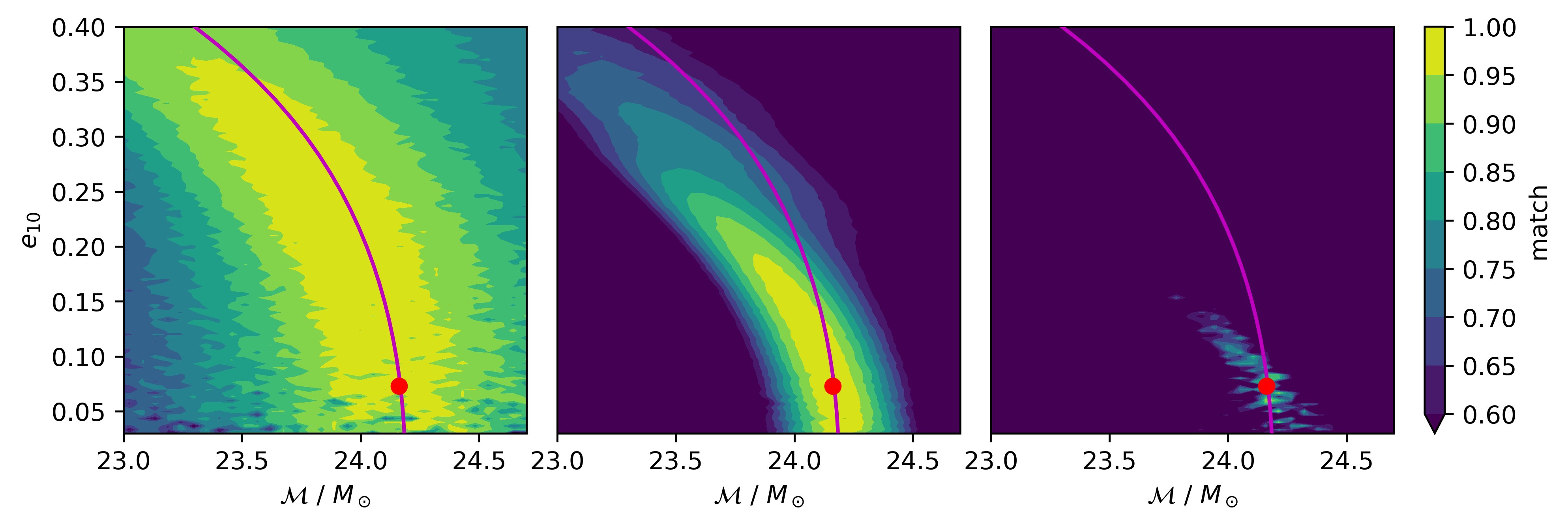}
\caption{Overlap between different eccentric waveform harmonics across the chirp mass and eccentricity space. at the time offset of the equivalent match for the $h_0$ waveform. The three plots show how the overlap of \textit{left:} $h_{-1}$ harmonic, \textit{centre:} $h_{1}$ harmonic and \textit{right} $h_{2}$ harmonic with the fiducial waveform varies. The overlap is calculated between a fixed waveform with $e_{10}=0.073$, $\mathcal{M}=\SI{24.14}{\Msolar}$, $q=2$ and zero component spins (indicated by the red dot and chosen such that our illustrative event lies on the degeneracy line) and the point shown. The waveform is generated from $\SI{10}{\hertz}$ and the overlap performed using a low frequency cutoff of $\SI{20}{\hertz}$. The time offset between waveforms is fixed by maximizing the overlap of the $h_{0}$ harmonics.  The magenta line shows the line of degeneracy between eccentricity and chirp mass for $h_{0}$ given in Eq.~(\ref{Eq:DegeneracyEqn}).  
}
\label{Fig:e_chirp_deg}
\end{figure*}

We would like to use the relative amplitudes of the eccentric harmonics to extract the binary eccentricity from a gravitational waveform.  In Sec.~\ref{Sec:deg_chirp}, we saw that the $k=0$ harmonic has a high match across a broad range of eccentricities, and derived a method of calculating the appropriate degenerate direction in $\mathcal{M}$ and $e_{10}$ space.  This opens up the possibility of using a single set of harmonics to probe the full extent of the degeneracy line, rather than searching across all values of mass and eccentricity independently. This approach is only viable if the overlap between mass and eccentricity follows a similar degeneracy for the other harmonics.  

In Fig.~\ref{Fig:e_chirp_deg}, we show the overlap with a fixed waveform for the $h_{-1}$, $h_{1}$ and $h_{2}$ harmonics across a range of chirp mass and eccentricity.  As discussed in Sec.~\ref{SubSec:EccHarms}, we are not free to vary the relative timing of the different harmonics.  Therefore, we first maximize the overlap for the $h_{0}$ harmonic and then use the same time delay when calculating the overlap for the other harmonics.  All three harmonics show a degeneracy between mass and eccentricity. For comparison, we show the degenerate direction for the $h_{0}$ waveform overlaid on the plots. While all three harmonics have a degeneracy between mass and eccentricity, they each follow a different degeneracy direction than $h_{0}$. This is unsurprising, given that each harmonic has a different frequency, $f_{k} = 2 f_{\phi} + k f_{\mathrm{r}}$.  Therefore, the $h_{1}$ and $h_{2}$ waveforms are at higher frequencies than $h_{0}$ at any instant while $h_{-1}$ is at a lower frequency. Furthermore, we know that the eccentricity of the orbit decays over time.  Thus, for a fixed value of $e_{10}$ the signal averaged eccentricity will be higher for $h_{2}$ and $h_{1}$ than it is for $h_{0}$ and lower for $h_{-1}$.  Thus, we expect the terms in the numerator of Eq.~(\ref{Eq:delta_m}) to be larger for $h_{1}$ and $h_{2}$ and therefore require a larger change in $\mathcal{M}$ for a fixed value of $e_{10}$.  This is indeed what we see, in particular for $h_{2}$ where the degeneracy between $e$ and $\mathcal{M}$ has a much lower slope.  We expect to observe the opposite effect for $h_{-1}$, but the degeneracy is remarkably similar to that for $h_{0}$, likely because the early part of the $h_{-1}$ signal is out of band.

Moving along the $h_{0}$ degeneracy, the overlap of $h_{1}$ remains above $0.9$ for eccentricities as high as $e_{10} \approx 0.2$, and for $h_{-1}$ up to $e_{10} \approx 0.4$.  Thus, even though these harmonics follow a different mass-eccentricity degeneracy, a single set of waveforms can be used for $h_{0}$, $h_{1}$ and $h_{-1}$ for eccentricities between $e_{10} = 0$ and at least $e_{10} = 0.2$.  Above that, the overlap of the $h_{1}$ waveform falls off rapidly.  Unfortunately, the degenerate direction for $h_{2}$ differs significantly from $h_{0}$.  Therefore, in what follows, we restrict attention to the $h_{1}$ and $h_{-1}$ harmonics, even though $h_{2}$ can contain as much \gls{snr} as $h_{-1}$ at moderate eccentricities (as shown in Fig.~\ref{Fig:harm_power}).  

In Appendix \ref{App:MassVariation} we show the variation of the match/overlap for the $h_{0}$, $h_{-1}$ and $h_{1}$ harmonics for binaries with $\mathcal{M} = \SI{10}{\Msolar}$ and $\SI{40}{\Msolar}$.  While the details differ, similar conclusions hold, namely that the degeneracy in the mass--eccentricity plane is similar enough in these three harmonics that a single set of waveforms can be used to cover a range of eccentricity values.

\section{Inferring the Eccentricity of a binary}
\label{Sec:Ecc_dist}

We have now introduced the key concepts required to enable rapid identification of eccentric systems. In Sec.~\ref{Sec:EccWaveform}, we presented the decomposition of the waveform into harmonics with frequencies of $2f_\phi + k f_{\mathrm{r}}$.  In Sec.~\ref{Sec:Decomposition}, we showed that the majority of the power in the signal, at least at moderate eccentricities, is contained in the first few harmonics and that the amplitudes of these sub-leading harmonics increases with eccentricity.  Finally, in Sec.~\ref{Sec:Degeneracy}, we showed that there is a mass--eccentricity degeneracy which allows a single set of eccentric harmonics to be used over a range of the parameter space.  Taken together, these features enable us to use a small set of waveforms to rapidly identify eccentric systems and provide an estimate of the eccentricity.  The method has clear parallels with similar proposals for identifying precession \cite{Fairhurst:2019vut} and higher \gls{gw} multipoles \cite{Mills:2020thr, Fairhurst:2023idl}.

In this section, we present a method to infer the eccentricity of a binary from the observed \gls{gw} signal.
We assume that the signal has previously been identified in the data and the best-fit parameters corresponding to a circular binary have been estimated through standard methods \cite{Veitch:2014wba, Ashton:2018jfp, Zackay:2018qdy, Fairhurst:2023idl}.  For the simplified example presented here, we focus only on the chirp mass and eccentricity.%
\footnote{The method should extend in a straightforward manner to the full parameter space of masses and aligned spins.  However, this will require an investigation of the degeneracy between eccentricity and other mass and spin parameters, such as mass ratio and effective spin, similar to that performed for chirp mass in Sec.~\ref{Sec:Degeneracy}.  That work is beyond the scope of this paper, but we plan to return to it in future.}  
We then use this information to generate a set of eccentric waveforms, obtain the eccentric harmonics and use them to probe the eccentricity of the system.  

\subsection{The example signal}
\label{Sec:example}

Let us consider an illustrative example of a binary with $\mathcal{M} = \SI{24.0}{\Msolar}$, $e_{10}=0.2$, $q=0.5$ and non-spinning components, with a \gls{snr} of $\rho \approx 20$.  We assume that the signal is identified by search and parameter estimation routines restricted to circular binaries that estimates a chirp mass of $\mathcal{M} = \SI{24.16}{\Msolar}$, which lies on the correct mass--eccentricity degeneracy line. We calculate the degeneracy in eccentricity and chirp mass as described in Sec.~\ref{Sec:Degeneracy} and generate the $k=0, 1, -1$ eccentric harmonics at a given point on the degeneracy line, using the method described in Sec.~\ref{SubSec:EccHarms}. As can be seen in Figure \ref{Fig:e_chirp_deg}, the degeneracy direction for the $h_{1}$ and $h_{-1}$ harmonics differs from $h_{0}$ and this determines the appropriate point: using a larger value of eccentricity will enable us to probe higher eccentricities, but at the expense of poorer matches at low eccentricity. From investigations, we have found that $e_{10} = 0.035$ provides good results for a binary with $\mathcal{M} = \SI{10}{\Msolar}$, and we extend to other masses through the scaling of $e_\mathrm{ref} \propto \mathcal{M}^{5/6}$, as discussed in Sec.~\ref{Sec:deg_chirp}.  Therefore, for this example, we use $e_{10} = 0.073$ when generating the eccentric harmonics.

We matched filter the $k=0, 1, -1$ eccentric harmonics against the data and use the ratio of \glspl{snr} in the different harmonics to infer the eccentricity. For our example, we obtain
\begin{align}\label{Eq:ecc_snr}
\rho_{0} &= 21.13 
, & 
\rho_{1} &= 4.04 \, ,\nonumber \\
\rho_{-1} &= 1.06 
, &
\rho_{(1, -1)} & = 4.11 \, , 
\end{align}
where $\rho_{(1, -1)}$ is the quadrature sum of the \glspl{snr} in the $k=1$ and $k=-1$ harmonics. 

\subsection{Variation of harmonic amplitudes with eccentricity}
\label{Sec:amp_variation}

In Fig.~\ref{Fig:harm_power}, we have seen that the \gls{snr} in the $k=1$ and $-1$ harmonics scales approximately linearly with the total \gls{snr}, although the exact value depends upon the initial location of the binary on its eccentric orbit. Furthermore, we are now working with a \textit{fixed} set of eccentric harmonics, computed at a fiducial eccentricity of $e_{10} = 0.073$. While these match well with waveforms of different eccentricity, see Fig.~\ref{Fig:match_e_mchirp} and Fig.~\ref{Fig:e_chirp_deg}, there is a reduction in the expected \gls{snr} due to mismatches between the fiducial waveforms and the signal.  In order to provide a mapping between the \glspl{snr} in each harmonic and the eccentricity, we must account for these effects.

\begin{figure}[t]
\centering
\includegraphics[width=\linewidth, valign=b]{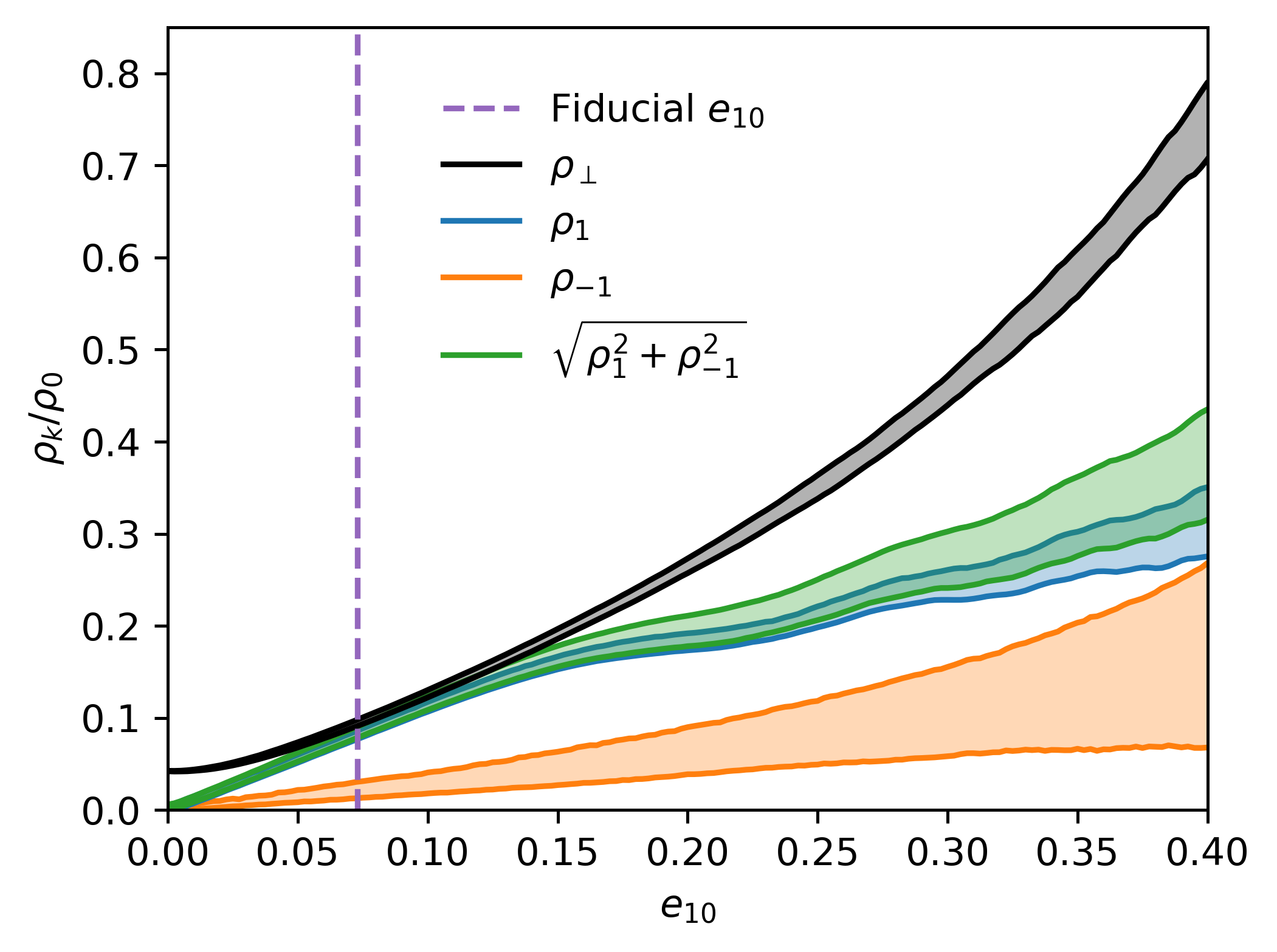}
\caption{The fractional \gls{snr} in different eccentric harmonics relative to the $h_{0}$ waveform.  The harmonics are generated at a fixed eccentricity of $e_{10}=0.073$ (dashed purple line) and mass, $\mathcal{M}=\SI{24.14}{\Msolar}$, using the methods described in Sec.~ \ref{Sec:Decomposition}.  Shaded regions correspond to the range of fractional \glspl{snr} in the different harmonics for binaries which lie along the mass-eccentricity degeneracy shown in Fig.~\ref{Fig:match_e_mchirp}. 
The  black region here shows the total power orthogonal to $h_{0}$, $\rho_{\perp} = \sqrt{\rho_\mathrm{total}^{2} - \rho_0^2}$/$\rho_0$. To determine the minimum and maximum values of each region we generate and match 32 \texttt{TEOBResumS-Dali} waveforms with varying initial mean anomaly for each eccentricity value.}
\label{Fig:min_max_lines}
\end{figure}

In Fig.~\ref{Fig:min_max_lines} we show the expected ratio of \glspl{snr} in the eccentric harmonics as a function of $e_{10}$, along the mass-eccentricity degeneracy line.  We have generated the harmonics at the fiducial eccentricity of $e_{10} = 0.073$. At each value of $e_{10}$ we generate a set of waveforms with different initial mean anomalies and calculate the match between each waveform and the $k=0$ harmonic, and the overlap with the other harmonics. This provides the expected value of $\rho_{k}/\rho_{0}$ for each waveform, which we show as a band on the figure. The relative \gls{snr} in the $h_{1}$ harmonic increases linearly up to $e_{10} \approx 0.15$ above which the rate of increase reduces, while the \gls{snr} in $h_{-1}$ increases approximately linearly, although with a large width, up to $e_{10} = 0.4$.  We also show the fractional power in the $h_{1}$ and $h_{-1}$ waveforms combined.%
\footnote{Given the large variation in expected \gls{snr} in the $h_{-1}$ harmonic, it is initially surprising that combining its contribution with $h_{1}$ leads to such an improvement in the expected \gls{snr} at high eccentricity.  We have investigated this effect and found that relative power in the two modes is anti-correlated, so that when there is less power in $h_{1}$ we have the maximum contribution from $h_{-1}$.  This explains the observed improvement, although we do not have a clear, physical interpretation of why this occurs.}
This increases with eccentricity, but the growth falls off similarly to the \gls{snr} in $h_{1}$.\textbf{}

On the figure, we also show the \gls{snr}, $\rho_{\perp}$, which is orthogonal to the $h_{0}$ waveform.  As expected, this increases with eccentricity.  For moderate eccentricities, the $h_{1}$ and $h_{-1}$ harmonics capture the vast majority of the orthogonal power.  However, at larger eccentricity, there is significant \gls{snr} which is not captured by the $h_{1}$ and $h_{-1}$ waveforms.  There are two reasons for this. First, at higher eccentricities, the other harmonics, most notably $h_{2}$, become more important and contribute a greater fraction of the \gls{snr}, as seen in Fig.~\ref{Fig:harm_power}.  Second, the match between the fiducial $h_{1}$ and $h_{-1}$ waveforms and those of the signal decreases, as shown in Fig.~\ref{Fig:e_chirp_deg}.  Both of these effects lead to a reduction in the fraction of available \gls{snr} that is recovered.  At very low eccentricities, we also see that there is a small amount of power orthogonal to $h_{0}$. Here, the impact of eccentricity is minimal, and the orthogonal \gls{snr} arises from a difference between the fiducial $h_{0}$ and the non-eccentric waveform.

\subsection{Consistency of $h_{1}$ and $h_{-1}$ signals}
\label{SubSec:PhaseConsistency}

Figure \ref{Fig:min_max_lines} provides the data required to map between the observed \gls{snr} in the $k = 0, 1$ and $-1$ eccentric harmonics and the eccentricity.  In many cases the observed \gls{snr} in the $h_{1}$ and $h_{-1}$ harmonics will be low and it is important to assess the uncertainties in the measured \glspl{snr} due to the presence of noise. The simplest approach is to add the \glspl{snr} in quadrature. While the quadrature sum of \glspl{snr} in the two modes contains all of the signal power, it also contains four noise components --- the real and imaginary parts of the matched filter of each harmonic.  We know from Eq.~(\ref{Eq:ecc_amps}) that both the amplitude and phase of these waveform components are correlated and, by making use of this, we can reduce the noise contribution to the combined \gls{snr}.  This will enable us to place tighter bounds on eccentricity. 

For simplicity, we assume that the other parameters of the system have been measured from the $h_{0}$ waveform and focus on determining the eccentricity and initial mean anomaly $l_{\mathrm{ref}}$ from the $h_{1}$ and $h_{-1}$ signals.  The initial phase $\Phi_{k}$ of the different harmonics in a signal is obtained from Eq.~(\ref{Eq:hp}) and Eq.~(\ref{Eq:hc}) as
\begin{equation}
    \Phi_{k} = 2 \gamma_\mathrm{ref} + (2+k) l_\mathrm{ref} \, ,
\end{equation}
so that
\begin{equation}\label{Eq:phase_consistency}
    \Phi_{1} - \Phi_{0} = \Phi_{0} - \Phi_{-1} = l_\mathrm{ref}\, .
\end{equation}

We wish to maximize the \gls{snr} over the amplitudes of the $h_{1}$ and $h_{-1}$ harmonics independently, but require a consistent phase between the three modes. To obtain the desired form of the \gls{snr}, we maximize the log-likelihood over the free amplitudes and phases.  The log-likelihood is given by \cite{Jaranowski:2005hz}
\begin{equation}
\log{\Lambda} = \mathrm{Re}(h|d) - \frac{1}{2} (h|h) = \frac{1}{2} \rho^2\, .
\end{equation}
Here, $h$ is the trial waveform given by
\begin{equation}
h = \sum_k A_k e^{i\Phi_k} h_k,
\end{equation}
and $d$ is the data.  We assume the signal has been identified and the parameters associated to the $h_{0}$ harmonic have been determined.  Then, we are only interested in determining the contributions of the $h_{1}$ and $h_{-1}$ eccentric harmonics to the signal.  

Let us denote the inner product between the data and waveform harmonics as
\begin{equation}
    (h_{k}|d) = \alpha_{k} e^{-i\varphi_{k}} \, .
\end{equation}
Then, we can rewrite the likelihood as
\begin{equation}\label{Eq:ecc_like}
\log{\Lambda}= \sum_{k \in [-1, 1]} \alpha_k A_k \cos(\Phi_k - \varphi_k) - \frac{1}{2} A_k^2 \, ,
\end{equation}
where we have used the orthonormality of the harmonics, $(h_j|h_k) = \delta_{jk}$.  If we maximize Eq.~(\ref{Eq:ecc_like}) independently over $A_{1, -1}$ and $\Phi_{1, -1}$ we obtain the quadrature sum of \glspl{snr} of the two harmonics, as expected.

We can enforce phase consistency of the harmonics by requiring that Eq.~(\ref{Eq:phase_consistency}) is satisfied.  Then the likelihood is
\begin{equation}\label{Eq:ecc_like_consist}
\log{\Lambda}= \sum_{k \in [-1, 1]} \alpha_k A_k \cos(kl_\mathrm{ref} + (\Phi_{0} - \varphi_k)) - \frac{1}{2} A_k^2 \, .
\end{equation}
We assume that $\Phi_{0}$ has been determined from the $h_{0}$ waveform and maximize over three parameters: $A_{1, -1}$ and $l_\mathrm{ref}$.  We first maximize over the amplitudes $A_{k}$ to obtain
\begin{equation}
    \hat{A}_{k} = \alpha_{k} \cos(kl_\mathrm{ref} + (\Phi_{0} - \varphi_k)) \, .
\end{equation}
Substituting the form of $\hat{A}_{k}$ into the likelihood, and re-expressing the cosine terms, we obtain
\begin{align}\label{Eq:ecc_like_amax}
\log{\Lambda}  = \sum_{k \in [-1, 1]} \frac{\alpha_k^{2}}{4} 
& \left[ 1  + \cos 2 l_\mathrm{ref} \cos(2 (\Phi_{0} - \varphi_k)) \right.
\nonumber \\
& \quad \left. - k \sin 2 l_\mathrm{ref} \sin(2 (\Phi_{0} - \varphi_k)) \right].
\end{align}
Next, we can maximize over $l_\mathrm{ref}$ to obtain an expression for $\tan{2 \hat{l}_{\mathrm{ref}}}$ which we substitute into Eq.~(\ref{Eq:ecc_like_amax}) to obtain
\begin{align}\label{Eq:ecc_max_like}
\log{\Lambda} &= \frac{1}{4}\left[ \alpha_{1}^{2} + \alpha_{-1}^{2} + 
\right. \\
& \left. \quad
\sqrt{\alpha_{1}^{4} + 2 \alpha_{1}^{2} \alpha_{-1}^{2} \cos[4\Phi_{0} - 2(\varphi_{1} + \varphi_{-1})] + \alpha_{-1}^4} \right]
\, .
\nonumber
\end{align}
We can check the expression in a couple of limiting cases.  First, if the phases of the two harmonics are consistent with a signal, then  $2\Phi_{0} - (\varphi_{1} + \varphi_{-1}) = 0$ and the log-likelihood simplifies to the usual form
\begin{equation}\label{Eq:ecc_in_phase}
\log{\Lambda} = \frac{1}{2}\left[ \alpha_{1}^{2} + \alpha_{-1}^{2}\right].
\end{equation}
Next, when the two harmonics are $90^{\circ}$ out of phase, $2\Phi_{0} - (\varphi_{1} + \varphi_{-1}) = \pi/2$, the likelihood becomes
\begin{align}\label{Eq:ecc_out_phase}
\log{\Lambda} &= \frac{1}{4}\left[ \alpha_{1}^{2} + \alpha_{-1}^{2}\right] + \frac{1}{4}| \alpha_{1}^{2} - \alpha_{-1}^{2}| 
\nonumber \\
& = \frac{1}{2} \mathrm{Max}(\alpha_{1}^{2}, \alpha_{-1}^{2}) \, .
\end{align}
Here, only the most significant harmonic contributes to the \gls{snr}.  Indeed, for phase offsets greater than $90^{\circ}$, we also instead obtain the result given in Eq.~(\ref{Eq:ecc_out_phase}) \footnote{Naively, it appears that when the components are maximally out of phase, $2\Phi_{0} - (\varphi_{1} + \varphi_{-1}) = \pi$, we again obtain the maximum \gls{snr} in Eq.~(\ref{Eq:ecc_max_like}). However, this requires one of the amplitudes $A_{k}$ to be negative, which is not permitted. Physically the likelihood cannot become higher when the two components become more out of phase with one another.  Nonetheless, it is always possible to simply take the \gls{snr} of the loudest harmonic as in Eq.~(\ref{Eq:ecc_out_phase}) by setting one of the $A_{k} = 0$.  Thus, the likelihood must therefore take the form given in Eq.~(\ref{Eq:ecc_out_phase}) whenever $2\Phi_{0} - (\varphi_{1} + \varphi_{-1}) > \pi/2$.}.

We denote the phase consistent combined \gls{snr} quantity as $\rho_{(1,-1)}$.   The benefit of the maximized \gls{snr} expression is that it reduces the number of noise degrees of freedom which contribute.  Heuristically, we expect to reduce the number of degrees of freedom from four to three, as we have eliminated one free phase.  Using the quadrature sum of $\rho_1$ and $\rho_{-1}$ gives four degrees of freedom, with the observed \gls{snr} described by a non-central $\chi^{2}$ distribution with four degrees of freedom. Enforcing phase consistency reduces the noise contribution. By generating $10^6$ simulated signals in different Gaussian noise realizations, we have seen that the distribution is almost identical to a non-central $\chi^{2}$ with three degrees of freedom.  Of course, if we require consistent phases of the harmonics, we must also require this of the signals, e.g. when constructing the relevant shaded region in Fig.~\ref{Fig:min_max_lines}.  We have done this and, as expected, there is negligible difference.  In principle, we are able to extract the initial mean anomaly from the relative phases of the harmonics.  Unfortunately, the value will depend sensitively upon the masses and eccentricity of the fiducial signal, as seen in \cite{Clarke:2022fma}, and therefore we are unlikely to significantly constrain its value.

In addition to requiring a consistent phases of the eccentric harmonics we could in principle also require the amplitudes to be consistent. However, due to the variation in \gls{snr} with mean anomaly, particularly for the $h_{-1}$ harmonic, this is not feasible.

\subsection{Estimating the Eccentricity from the harmonic amplitudes}

\begin{figure*}[ht]
\centering
\includegraphics[width=0.9\linewidth, valign=b]{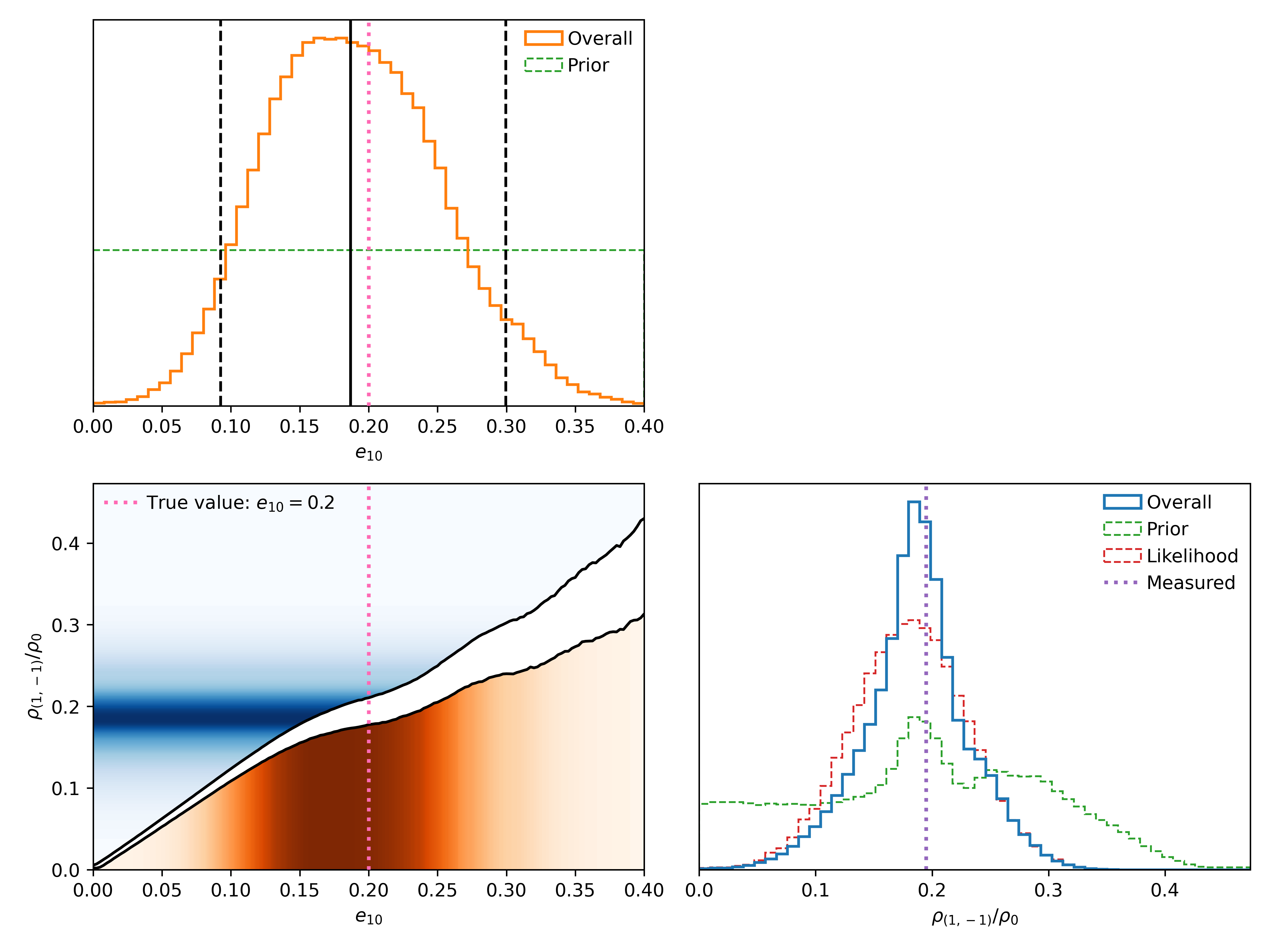}
\caption{
Mapping of measured \glspl{snr} to eccentricity for a single detector and Gaussian noise example with an simulated signal of $e_{10}=0.2, \mathcal{M}=\SI{24}{\Msolar}, q=0.5$ and a total \gls{snr} of 20. The bottom right panel shows the measured and inferred distribution of \gls{snr} ratio in the $h_{1}$ and $h_{-1}$ harmonics relative to $h_{0}$.  The prior on \gls{snr} ratio is obtained by requiring a flat prior on eccentricity, as discussed in the text.  The overall distribution is the product of the likelihood and prior. The bottom left panel shows the mapping between \gls{snr} ratio and eccentricity, along a chirp mass--eccentricity degeneracy line passing through $e_{10}=0.2, \mathcal{M}=\SI{24}{\Msolar}$.  The shaded regions show the distributions of the \gls{snr} ratio and eccentricity, with darker shades corresponding to greater probability density. The top left panel shows the inferred eccentricity distribution with the black lines corresponding to quantiles of 5\%, 50\%, and 95\%. The dotted pink line represents the simulated eccentricity, $e_{10} = 0.2$.}
\label{Fig:ecc_dist}
\end{figure*}

To obtain an estimate of the eccentricity, we calculate the relative \gls{snr} in $h_{1}$ and $h_{-1}$ and then use Fig.~\ref{Fig:min_max_lines} to read off the eccentricity. For our example, we have $\rho_{(1, -1)}/\rho_{0} = 4.11/21.13 \approx 0.2$ and, from Fig.~\ref{Fig:min_max_lines}, we obtain an eccentricity of $e_{10} \approx 0.2$.  We know, however, that the measured \gls{snr} is comprised of both a signal and noise contribution, this provides an uncertainty on the \gls{snr} in the eccentric signal, as opposed to the noise.  In addition, the mapping from \gls{snr} to eccentricity is complicated due to the width of the allowed region.  We must address both of these issues in order to obtain an accurate measure of the eccentricity, and its uncertainty, from the measured \glspl{snr}, as shown in Fig.~\ref{Fig:ecc_dist}.

It is straightforward to map from an expected value of signal \gls{snr} to the distribution of measured \glspl{snr} in noise using a non-central $\chi^2$ distribution \cite{Fairhurst:2023idl}. We wish to obtain the inverse mapping: from the measured \gls{snr} to a distribution for the expected signal \gls{snr}, which we achieve through rejection sampling.  First, we generate a uniform distribution of expected signal \gls{snr} values and for each find the corresponding probability density of obtaining the given measured \gls{snr} in noise. We then use these probabilities as weights to either accept or reject each signal \gls{snr} sample, with higher probabilities leading to a greater chance of acceptance.  These accepted samples then form the distribution of signal \gls{snr} for $\rho_{(1, -1)}$, which we divide by measured value of $\rho_{0}$ to get the \gls{snr} ratio. In principle, we should also account for noise in the $\rho_{0}$ measurement but this has a negligible effect due to the higher \gls{snr} of $h_{0}$ and we choose not to incorporate it.  The distribution of signal \gls{snr} ratios is plotted as the \textit{likelihood} in the bottom right panel of Fig.~\ref{Fig:ecc_dist}.  The signal \gls{snr} distribution peaks at a smaller value than the measured \gls{snr}, as is expected since noise is more likely to increase the measured \gls{snr}.

Next, we need to derive a mapping from the ratio of \glspl{snr} in the eccentric harmonics to the binary's eccentricity.  To do so, we must account both for the uncertainty in the mapping which arises from the variation of \gls{snr} with initial mean anomaly and the non-linearity of the mapping between parameters, as shown in the lower right panel in Fig.~\ref{Fig:ecc_dist}.  We would like to enforce a uniform prior on eccentricity.%
\footnote{Astrophysically, there may be an argument to impose different eccentricity priors.  This is straightforward to do starting with a uniform prior on $e_{10}$.  In contrast, there is no (astro-)physical motivation to apply a specific prior on the \gls{snr} ratio.}
We obtain the prior on the \gls{snr} ratio, $\rho_{(1, -1)}/\rho_{0}$ numerically by generating $10^6$ samples from the uniform prior on eccentricity and mapping them to \gls{snr} ratio.  Each value of $e_{10}$ corresponds to a range of permitted \gls{snr} ratios and we randomly select a value uniformly distributed within the permitted range.  A uniform prior in eccentricity maps to the prior distribution shown in the bottom right panel of Fig.~\ref{Fig:ecc_dist}.  The features at \gls{snr} ratio around $e_{10}=0.2$ arise from the structure in the mapping to eccentricity.  The falloff at $e_{10}=0.4$ is due to our truncation of the mapping at $e_{10} = 0.4$ but this is at large enough values that it does not impact our result.

Combining the prior and likelihood distributions gives the overall distribution for the \gls{snr} ratio, again shown in Fig.~\ref{Fig:ecc_dist}.  We map this distribution back to eccentricity to obtain the measured eccentricity distribution of the signal.

Figure \ref{Fig:ecc_dist} shows the final result. We have generated a gravitational wave signal from a binary with $\mathcal{M} = \SI{24}{\Msolar}$ on an eccentric orbit with $e_{10} = 0.2$, with an \gls{snr} of 20 injected into Gaussian noise.  By matched filtering the data, we obtained the \glspl{snr} in the eccentric harmonics given in Eq.~(\ref{Eq:ecc_snr}).  From these \glspl{snr}, we obtain the probability distribution for $\rho_{(1,-1)}/\rho_0$.  We then map this to an eccentricity distribution, and obtain the value
\begin{equation}
    e_{10} = 0.19^{+0.11}_{-0.10} \, ,
\end{equation}
where the central value is the median and the uncertainties provide the 90\% confidence interval. Most significantly, there is sufficient \gls{snr} in the $h_{1}$ and $h_{-1}$ eccentric harmonics to confidently identify the signal as originating from an eccentric binary and place a lower bound (at 90\% confidence) of $e_{10} = 0.09$.

\section{Discussion}
\label{Sec:Discussion}

We have investigated the structure of the gravitational wave signal emitted by an eccentric binary merger and seen that the waveform can be decomposed into a series of harmonics, as has been shown previously in the literature for the inspiral portion of the waveform \cite{Moreno:1995, Seto:2001pg, Willems:2007nq, Valsecchi_2012}. The frequencies of these harmonics are given by $2f_{\phi} + k f_{\mathrm{r}}$, where $f_{\mathrm{r}}$ is the radial frequency (characterised by the time taken to return to apoapsis) and $f_{\phi}$ is the azimuthal frequency (characterised by the time taken to return to a fixed direction) \cite{Moore:2016qxz}. With the use of an \gls{svd} of eccentric waveforms, we have shown that this decomposition also works well through the merger and ringdown, and provided a framework for efficiently constructing these harmonics from existing eccentric waveform models.

We have shown that there are three modes which, at low to moderate eccentricities,  contain the majority of the power in the \gls{gw} signal, which we denote as $h_{0}$, $h_{1}$, and $h_{-1}$. In all cases we have studied, $h_{0}$ represents the dominant  mode, containing the majority of the \gls{snr} and resembling the signal emitted from a non-eccentric binary of comparable mass.  The next most important is the $h_{1}$ waveform (first higher frequency) and subsequently $h_{-1}$ (first lower frequency) which describe the basic form of the modulations due to the binary orbiting between apo- and periapsis. The amplitudes of the sub-leading modes increase approximately linearly with eccentricity and so, if we are able to measure the relative amplitudes of these modes, we can extract the eccentricity of the binary.

Next, we have identified the degenerate direction in chirp mass and eccentricity space by adapting the method of \cite{Favata:2021vhw}. We have showed that while the degeneracy of the $h_{0}$ mode clearly follows the expected degeneracy line, the degeneracy line for the sub-leading harmonics differs slightly. Nonetheless, the degenerate direction for $h_{0}$, $h_{1}$ and $h_{-1}$ is similar enough that we can use a common direction determined by $h_{0}$ as a reasonable approximation of all three. This allows the three waveforms to probe a range of eccentricities at which these harmonics are a reasonably good representation of the waveform.

We have proposed a method to rapidly estimate the eccentricity of a binary merger with minimal computational cost.  Beginning from the best-fit non-eccentric parameters, we generate the $h_0$, $h_1$, and $h_{-1}$ eccentric harmonics and matched filter them against the data. The ratio of the SNRs in each harmonic should then give us information about the eccentricity of the binary. This is complicated somewhat by the fact that the amplitude of the different harmonics also varies with the initial mean anomaly of the system, which we handle by appropriately broadening the mapping from the harmonic \glspl{snr} to eccentricity.  At the end of Sec.~\ref{Sec:Ecc_dist}, we demonstrate the method on a simulated signal and show that we can accurately recover the eccentricity.

One of the benefits of our proposed approach is its speed.  In order to investigate a range of eccentricities, we obtain the three leading eccentric harmonics by generating six waveforms.  We then filter these three harmonics against the data and, from the results, are able to quantify the eccentricity of the signal.  In translating from observed \glspl{snr} to eccentricity, we require a mapping that accounts for variation with initial mean anomaly, as shown in Fig.~\ref{Fig:min_max_lines}.  Producing that requires a significant number of overlaps between eccentric waveforms.  In particular, we must generate the harmonics and then calculate overlaps with signals with a range of eccentricities and anomalies.  This takes on the order of an hour of CPU time but in practice we are able to pre-compute these grids of overlaps for a discrete selection of masses, and interpolate the results to an arbitrary mass. This computation is only required once, and can then be used on all the signals from a given observing run. The time taken to analyze a single event is under a minute.

The work here provides a novel method for rapidly inferring eccentricity in a binary system.  However, there are several additional steps required before this can be used as a tool to apply to recent \gls{gw} observations.  First, we have restricted attention to the chirp mass--eccentricity space, holding both the mass ratio and effective spin constant.  While this is a reasonable starting point, particularly as the impact of eccentricity is most strongly degenerate with chirp mass, we must extend the formulation across the four dimensional space of masses, effective spin and eccentricity. This will require a detailed investigation of the degeneracies involved. Next, we must incorporate this eccentricity measurement into the full parameter estimation, rather than providing only a two-dimensional measure of mass and eccentricity.  The methods provided in the \texttt{simple-pe} formalism \cite{Fairhurst:2023idl} can be naturally extended to include eccentricity as it already contains similar methods applied to higher \gls{gw} multipoles \cite{Mills:2020thr} and precession \cite{Fairhurst:2019vut}. We also plan to investigate extending this method to signals with higher eccentricities by repeated decomposition and matched filtering at higher fiducial eccentricity values (especially useful for low chirp masses, see Appendix \ref{App:MassVariation}).

There are several other applications of the methods that introduced here. First, the harmonic decomposition of the eccentric waveform provides a possible basis for searches for eccentric binaries.  We have shown that the first three eccentric harmonics contain a large fraction of the signal power and, furthermore, a single set of waveform harmonics can cover a range of eccentricities. Therefore, replacing a single waveform with these three eccentric harmonics enables a more efficient search for eccentric binaries up to moderate values of eccentricity. This approach has previously been proposed \cite{Fairhurst:2019vut} and applied \cite{McIsaac:2023ijd} to precessing binaries, and higher \gls{gw} multipoles \cite{Wadekar:2024zdq}.

Existing methods of searching for eccentricity in binary mergers often start from the non-eccentric parameter estimates \cite{Romero-Shaw:2019itr, Romero-Shaw:2022xko} and then use these to infer eccentricity by ``unwrapping'' the additional dimensions.  Our analysis of
the appropriate direction to unwrap, and particularly the fact that the chirp mass should vary as the eccentricity changes will be of use to those analyses. 

Several papers have discussed the fact that, particularly for high mass signals, it can be difficult to distinguish the impact of orbital eccentricity from precession of the binary orbit \cite{Romero-Shaw:2022xko, Gupte:2024jfe, Romero-Shaw:2022fbf, Xu:2022zza}.  Our waveform decomposition into harmonics, along with the harmonic decomposition of the precessing waveform \cite{Fairhurst:2019vut} provide the ideal tools to investigate this degeneracy. For both systems, the leading order waveform looks essentially the same as a non-precessing, circular binary.  Then, we have the leading order corrections due to eccentricity and precession.  We can compare these leading order corrections.  Calculating the overlaps between precession and eccentricity effects will inform us about the mass threshold the two effects become degenerate, and may even provide a degeneracy mapping between precession and eccentricity in these degenerate regions. The studies will be complicated by the short waveforms at these high mass values, meaning the various harmonics are not orthogonal.

\section*{Acknowledgements}
We thank Christian Chapman-Bird, Teagan Clarke, Neil Cornish, Paul Lasky, Isobel Romero-Shaw, Eric Thrane, and Aditya Vijaykumar for useful discussions and suggestions. The authors are also grateful for computational resources provided by LIGO Laboratory and supported by National Science Foundation grants PHY-0757058 and PHY-0823459 and provided by Cardiff University, and funded by STFC grant ST/I006285/1.  SF thanks the Science and Technology Facilities Council (STFC) for support through grant ST/V005618/1. BP thanks the STFC for support through grant ST/Y509152/1. Plots were prepared with \texttt{Matplotlib}~\cite{Hunter:2007} with analysis making use of {\texttt{NumPy}}~\cite{harris2020array}, 
\texttt{pycbc}~\cite{pycbc-software}, and {\texttt{Scipy}}~\cite{2020SciPy-NMeth}.

\bibliography{references}

\appendix

\section{Variation with mass}
\label{App:MassVariation}

\begin{figure*}[tbh]
\centering
\includegraphics[width=\linewidth, valign=b]{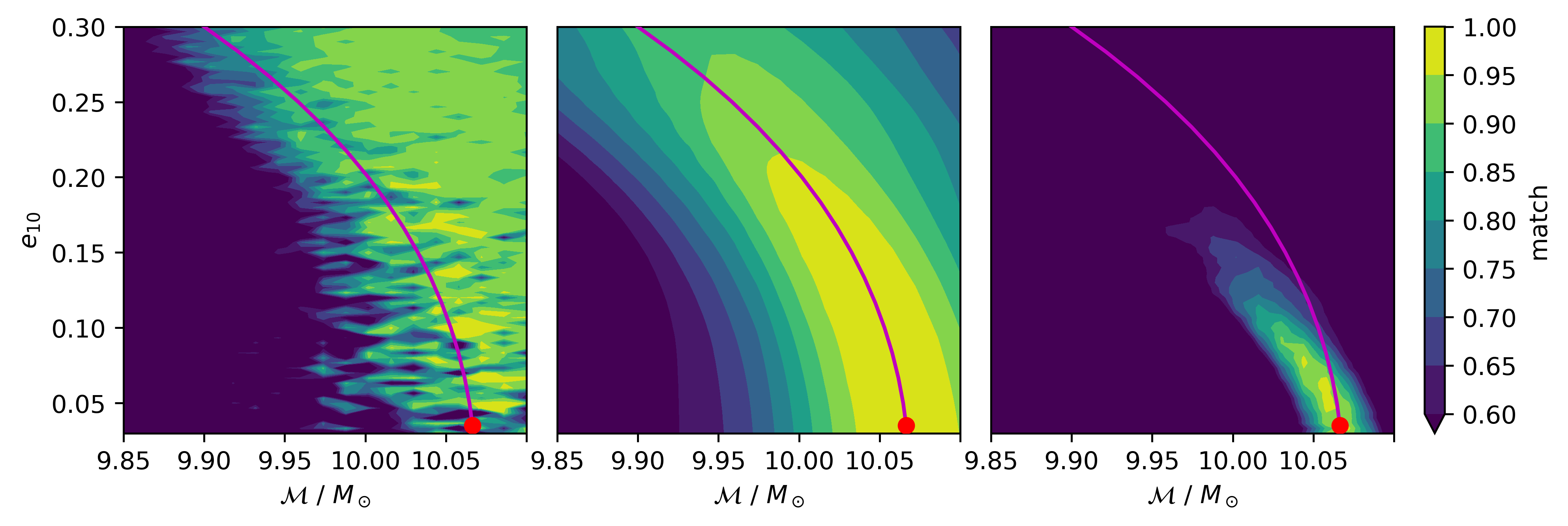}
\caption{Overlap between different eccentric waveform harmonics across the chirp mass and eccentricity space at the time offset of the equivalent match for the $h_0$ waveform. The three plots show how the match varies for the $h_{-1}$ harmonic (left panel), $h_{0}$ harmonic (centre panel) and $h_{1}$ harmonic (right panel).  In all cases, the match is calculated between the fiducial waveform, generated at $e_{10}=0.035, \mathcal{M}=\SI{10.07}{\Msolar}$ (indicated by the red dot) and the point shown. In all cases the system has a mass ratio of $q=0.5$, the waveform is generated from $\SI{10}{\hertz}$ and the match performed using a low frequency cutoff of $\SI{20}{\hertz}$. The magenta line shows the line of degeneracy between eccentricity and chirp mass described by Eq.~(\ref{Eq:DegeneracyEqn}).
}
\label{Fig:match_degeneracy_10msun}
\end{figure*}

\begin{figure*}[tbh]
\centering
\includegraphics[width=\linewidth, valign=b]{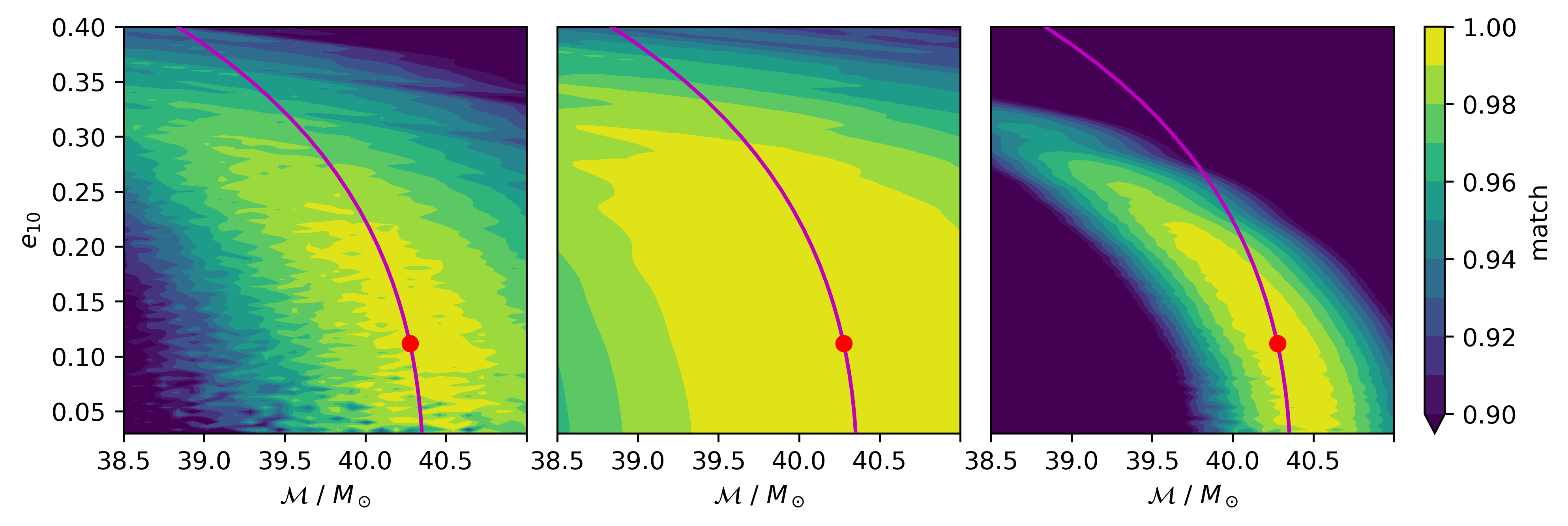}
\caption{Overlap between different eccentric waveform harmonics across the chirp mass and eccentricity space at the time offset of the equivalent match for the $h_0$ waveform. The three plots show how the match varies for the $h_{-1}$ harmonic (left panel), $h_{0}$ harmonic (centre panel) and $h_{1}$ harmonic (right panel).  In all cases, the match is calculated between the fiducial waveform, generated at $e_{10}=0.112, \mathcal{M}=\SI{40.28}{\Msolar}$ (indicated by the red dot) and the point shown. In all cases the system has a mass ratio of $q=0.5$, the waveform is generated from $\SI{10}{\hertz}$ and the match performed using a low frequency cutoff of $\SI{20}{\hertz}$. The magenta line shows the line of degeneracy between eccentricity and chirp mass described by Eq.~(\ref{Eq:DegeneracyEqn}).  
}
\label{Fig:match_degeneracy_40msun}
\end{figure*}

In the main body of the paper, we have focused on the analysis of a binary with $\mathcal{M} = \SI{24}{\Msolar}$ and mass ratio $q=0.5$. In this Appendix, we investigate the applicability of the methods presented in the body of the paper across the mass space.  
For this, we require that the mass--eccentricity degeneracy matches the expected direction, as calculated in Section \ref{Sec:Degeneracy} and, equally importantly, that the degeneracy in match of the sub-leading harmonics and in particular $h_{1}$ and $h_{-1}$ is along a similar direction to the leading $h_{0}$ harmonic.
In Fig.~\ref{Fig:match_degeneracy_10msun} and Fig.~\ref{Fig:match_degeneracy_40msun} we show the match across mass and eccentricity parameter space with again a binary with $e_{10} = 0.2$, $\mathcal{M}=[\SI{10}{\Msolar}, \SI{40}{\Msolar}]$ and $q = 0.5$.  In both cases, the degeneracy is along the expected direction (as calculated using Eq.~(\ref{Eq:DegeneracyEqn})) and extends to $e_{10} \approx 0.2$ or $0.3$ at $\mathcal{M} = \SI{10}{\Msolar}$ and $\mathcal{M} = \SI{40}{\Msolar}$ respectively. The degeneracy of the $h_{-1}$ harmonic is consistent with the the $h_{0}$ direction.  However, the $h_{1}$ has a different degeneracy direction and the match decays more rapidly.  This limits the domain of applicability of our decomposition, particularly for the $\mathcal{M} = \SI{10}{\Msolar}$ signal.

\begin{figure*}[tbh]
\centering
\includegraphics[width=0.49\linewidth, valign=b]{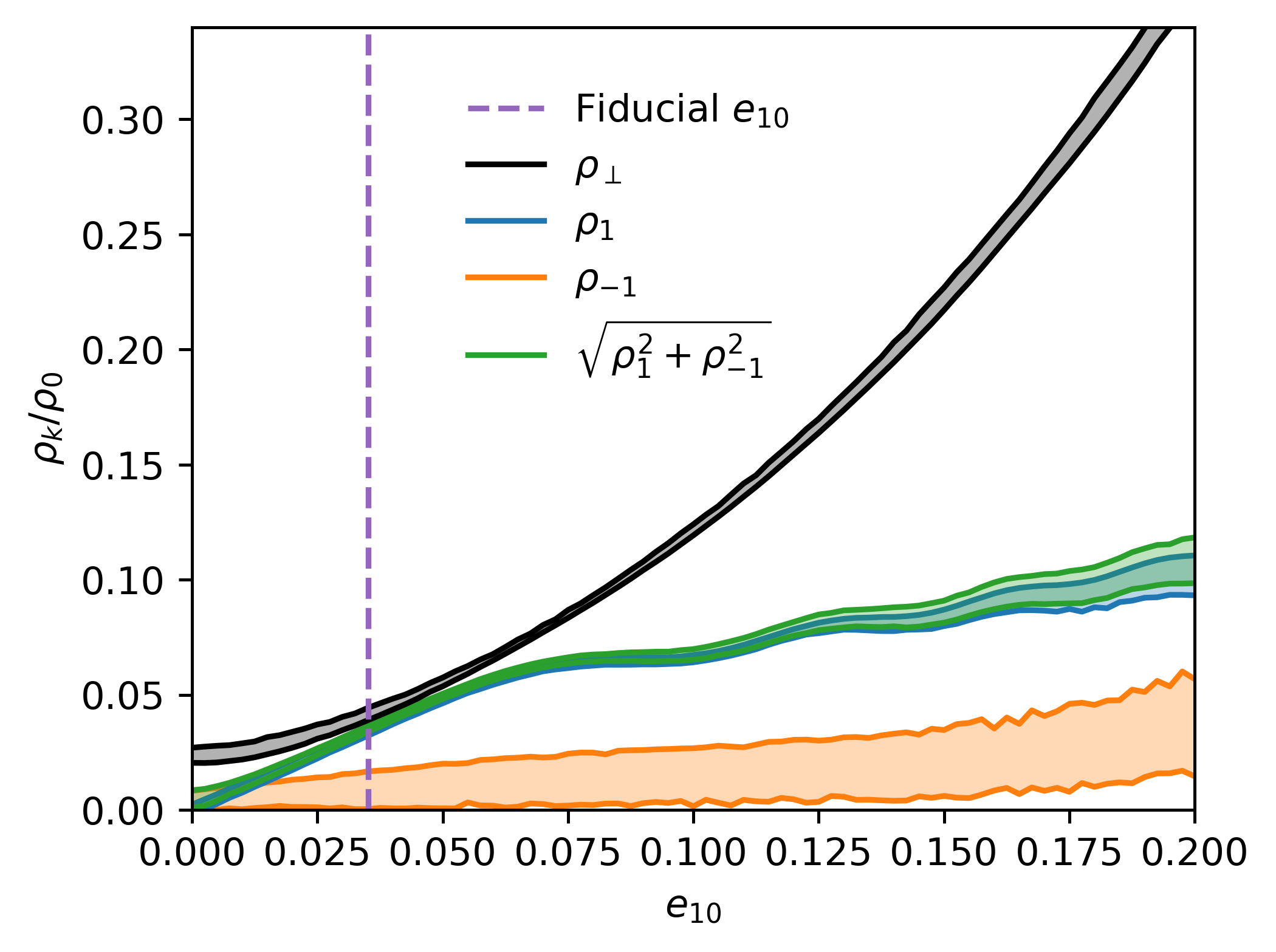}
\includegraphics[width=0.49\linewidth, valign=b]{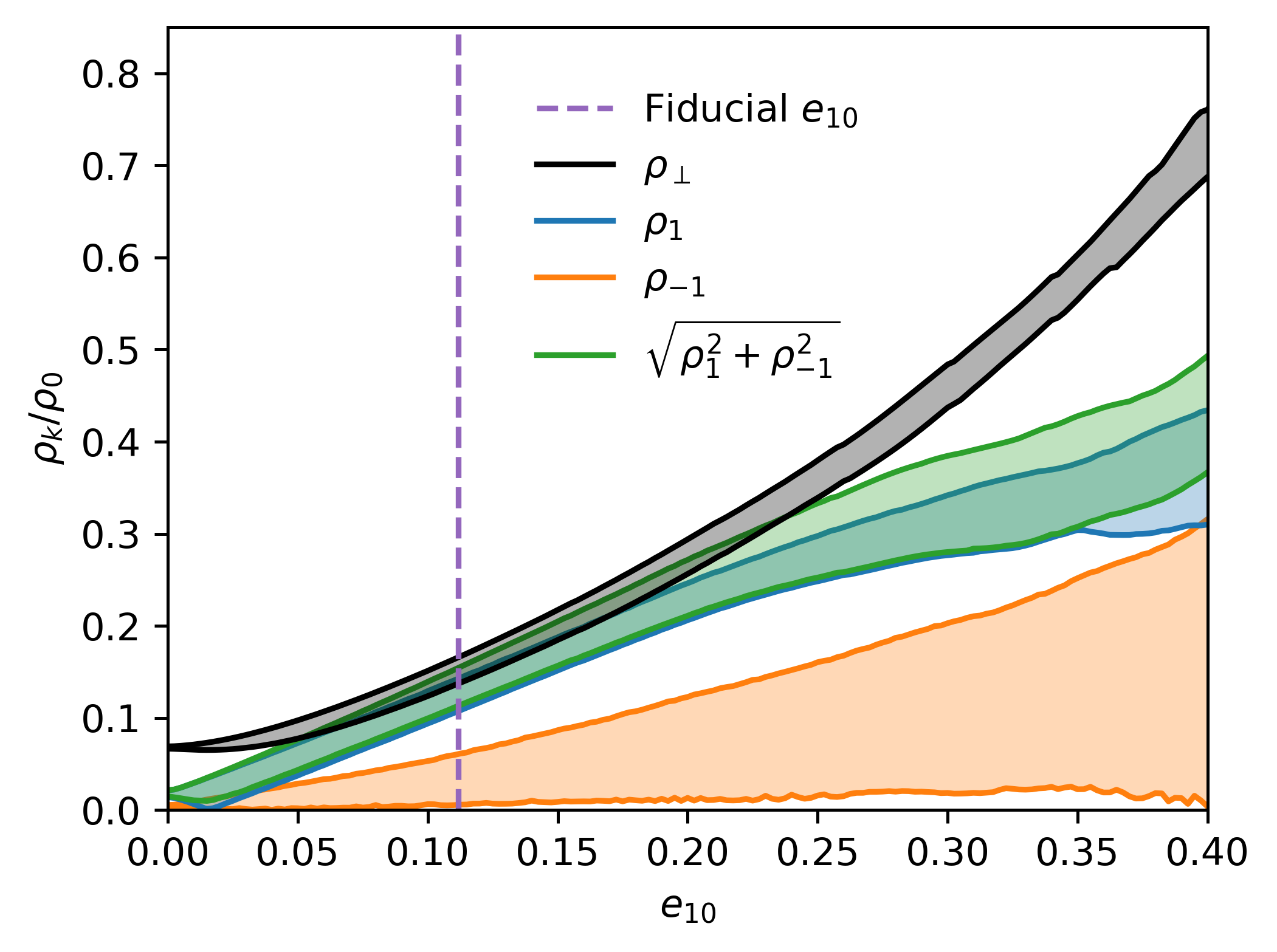}
\caption{The shaded regions correspond to the range of possible matches between different harmonics at fiducial values of $e_{10}=0.035$, $\mathcal{M}=\SI{10.07}{\Msolar}$ (left panel) or $e_{10}=0.112$, $\mathcal{M}=\SI{40.28}{\Msolar}$ (right panel), with a set of trial waveforms at different eccentricities along the corresponding degeneracy line between chirp mass and eccentricity (\ref{Eq:DegeneracyEqn}), all divided by the equivalent match with $h_0$. The blue and orange regions show $\rho_1$/$\rho_0$ and $\rho_{-1}$/$\rho_0$ respectively, the green region shows the region corresponding to $\sqrt{\rho_1^2+\rho_{-1}^2}$/$\rho_0$, and the black line shows the total power in the higher harmonics, calculated as $\sqrt{\rho_\mathrm{total}^{2} -\rho_0^2}$/$\rho_0$. To determine the minimum and maximum values of each region we generate and match 32 \texttt{TEOBResumS} waveforms with equally spaced apsidal anomaly values between 0 and $2\pi$ for each eccentricity value.}
\label{Fig:min_max_different_mass}
\end{figure*}

We also require that the amplitude of the $h_{1}$ and $h_{-1}$ harmonics, obtained at the fiducial value of $e_{10}$, can be used to infer the eccentricity.  For this, we require that the ratio of \glspl{snr} in the sub-leading harmonics, relative to $h_{0}$, increases as a function of eccentricity. Figure \ref{Fig:min_max_different_mass} shows the variation of the \gls{snr} ratios with eccentricity. For the $\mathcal{M}=\SI{10}{\Msolar}$ system the \glspl{snr} in the $h_{1}$ and $h_{-1}$ harmonics increases up to $e_{10} \approx 0.075$. Above this eccentricity there is significant power orthogonal to the three leading harmonics, which limits the efficacy of our approach.  For higher eccentricities the decomposition at $e_{10} = 0.0035$ and restriction to three harmonics is not sufficient to accurately describe the waveforms.  Based on Fig.~\ref{Fig:min_max_different_mass}, the dominant effect is likely the mismatch between $h_{1}$ generated at our fiducial point and the features in the waveform.  It is possible that recalculating the harmonics at higher values of $e_{10}$ will enable us to extend the validity of this approach.  For the high mass system, the \gls{snr} ratio between the $h_{1, -1}$ and $h_{0}$ harmonics increases up to $e_{10}\gtrsim 0.4$ and captures the vast majority of the signal power orthogonal to $h_{0}$ up to $e_{10} \sim 0.3$.
 
Combined, the results of Figs.~\ref{Fig:match_degeneracy_10msun}, \ref{Fig:match_degeneracy_40msun} and \ref{Fig:min_max_different_mass} provide good evidence that the proposed method is applicable over a range of masses. The range of eccentricities for which it is applicable will vary with mass, with a more restricted range of eccentricity for lower masses.

\end{document}